\documentclass[]{IEEEtran}
\usepackage{graphicx,epsfig,amsmath,amsbsy,amssymb,citesort,verbatim,url,algorithm,algpseudocode,soul}
\usepackage[usenames]{color}
\usepackage{setspace,amsfonts,epsf,url}
\usepackage{subfigure}
\usepackage{bbm}
\usepackage[normalem]{ulem}

\let\labelindent\relax
\usepackage{enumitem}
\setlist[itemize]{leftmargin=*}
\setlist[enumerate]{leftmargin=*}

\newcommand{\x}{{\bf x}}
\newcommand{\A}{{\bf A}}
\newcommand{\y}{{\bf y}}
\newcommand{\z}{{\bf z}}
\newcommand{\f}{{\bf f}}
\newcommand{\w}{{\bf w}}
\newcommand{\n}{{\bf n}}

\newtheorem{myRemark}{Remark}

\newtheorem{myDef}{Definition}
\newtheorem{myLemma}{Lemma}
\newtheorem{myCoro}{Corollary}
\newtheorem{myTheorem}{Theorem}

\def \l {\left}
\def \r {\right}

\usepackage{etoolbox}
\newtoggle{TSP}
\togglefalse{TSP}

\begin{document}

\title{Optimal Trade-offs in Multi-Processor Approximate Message Passing}
\author{Junan Zhu,~\IEEEmembership{Student Member,~IEEE},
Dror~Baron,~\IEEEmembership{Senior Member,~IEEE}, and Ahmad Beirami,~\IEEEmembership{Member,~IEEE}
\thanks{Part of the work has appeared in Han et al.~\cite{HanZhuNiuBaron2016ICASSP} and Zhu et al.~\cite{ZhuBeiramiBaron2016ISIT}.}
\thanks{The work was supported by the
National Science Foundation under the Grants CCF-1217749 and ECCS-1611112.}
\thanks{Junan Zhu and Dror Baron are with the Department of Electrical and Computer Engineering, NC State University, Raleigh, NC 27695.
E-mail: \{jzhu9, barondror\}@ncsu.edu. Ahmad Beirami is with the Research Laboratory of Electronics, Massachusetts Institute of Technology. E-mail: beirami@mit.edu.}
}
\maketitle

\begin{abstract}
We consider large-scale linear inverse problems in Bayesian settings. We follow a recent line of work that applies the
approximate message passing (AMP) framework to multi-processor (MP)
computational systems, where each processor node stores and processes
a subset of rows of the measurement
matrix along with corresponding measurements.
In each MP-AMP iteration, nodes of the MP system and its fusion center
exchange lossily compressed messages pertaining to their estimates of the input.
In this setup, we derive the optimal per-iteration coding rates using dynamic programming.
We analyze the excess mean squared error (EMSE) beyond the minimum mean squared error (MMSE), and
prove that, in the limit of low EMSE,
the optimal coding rates increase approximately linearly per iteration. Additionally, we obtain that the combined cost of computation
and communication scales with the desired estimation quality according to $O(\log^2(1/\text{EMSE}))$.
Finally, we study trade-offs between the physical
costs of the estimation process including computation time,
communication loads, and the estimation quality as a multi-objective optimization problem,
and characterize the properties of the Pareto optimal surfaces.

\end{abstract}

\section{Introduction}
\subsection{Motivation}
Many scientific and engineering problems~\cite{DonohoCS,CandesRUP} can be approximated
as linear systems of the form
\begin{equation}
\y = \A\x + \z,
\label{eq:matrix_channel}
\end{equation}
where $\x\in\mathbb{R}^N$ is the unknown input signal, $\A\in\mathbb{R}^{M\times N}$ is the matrix that characterizes the linear system, and $\z\in\mathbb{R}^M$ is measurement noise.
The goal is to estimate $\x$ from the noisy measurements $\y$ given $\A$ and statistical information about $\z$; this is a {\em linear inverse problem}. Alternately, one could view the estimation of $\x$ as fitting or learning a linear model for the data comprised of $\y$ and $\A$.

When $M\ll N$, the setup~\eqref{eq:matrix_channel} is known as compressed sensing (CS)~\cite{DonohoCS,CandesRUP}; by posing a sparsity or compressibility
requirement on the signal,
it is indeed possible to accurately recover $\x$ from the ill-posed linear system~\cite{DonohoCS,CandesRUP} when the number of measurements $M$ is large enough, and the noise level is modest. However, we might need $M>N$ when the signal is dense or the noise is substantial. Hence, we do not constrain ourselves to the case of $M\ll N$.

Approximate message passing (AMP)~\cite{DMM2009,Montanari2012,Bayati2011,Rush_ISIT2016_arxiv} is an iterative framework that solves  linear inverse problems by successively decoupling~\cite{Tanaka2002,GuoVerdu2005,GuoWang2008} the problem in~\eqref{eq:matrix_channel} into scalar denoising
problems with additive white Gaussian noise (AWGN). AMP has received considerable attention, because of its fast convergence and the state evolution (SE) formalism~\cite{DMM2009,Bayati2011,Rush_ISIT2016_arxiv}, which offers a precise
characterization of the AWGN denoising problem in each iteration.
In the Bayesian setting, AMP often achieves the minimum mean squared error
(MMSE)~\cite{GuoBaronShamai2009,RFG2012,ZhuBaronCISS2013,Krzakala2012probabilistic} in the limit of large linear systems ($N\rightarrow\infty, \frac{M}{N}\rightarrow \kappa$).

In real-world applications, a multi-processor (MP) version of the linear system could be of interest, due to either storage limitations in each individual processor node, or the need for fast computation. This paper considers  multi-processor linear systems (MP-LS)~\cite{Mota2012,Patterson2014,Han2014,Ravazzi2015,Han2015SPARS,HanZhuNiuBaron2016ICASSP}, in which there are $P$ {\em processor nodes} and a {\em fusion center}. Each processor node stores $\frac{M}{P}$ rows of the matrix $\A$, and acquires the corresponding linear measurements of the underlying signal $\x$. Without loss of generality, we model the measurement system in processor node $p\in \{1,...,P\}$ as
 \begin{equation}\label{eq:one-node-meas}
    y_i=\A_i \x+z_i,\ i\in \left\{\frac{M(p-1)}{P}+1,...,\frac{Mp}{P}\right\},
 \end{equation}
 where $\A_i$ is the $i$-th row of $\A$, and $y_i$ and $z_i$ are the $i$-th entries of $\y$ and $\z$, respectively.
Once every $y_i$ is collected, we run distributed algorithms among the fusion center and $P$ processor nodes to estimate the signal $\x$.
MP versions of AMP (MP-AMP) for MP-LS have been studied in the literature~\cite{Han2014,HanZhuNiuBaron2016ICASSP}.
Usually, MP platforms are designed for distributed settings such as sensor networks~\cite{pottie2000,estrin2002} or large-scale ``big data" computing systems~\cite{EC2}, where the computational and communication burdens can differ among different settings. We reduce the communication costs of MP platforms by applying lossy compression~\cite{Berger71,Cover06,GershoGray1993} to the communication portion of MP-AMP.
Our key idea in this work is to minimize the total communication and computation costs by varying the lossy compression schemes in different iterations of MP-AMP.

\subsection{Contribution and organization}
Rate-distortion (RD) theory suggests that we can transmit data with greatly reduced coding rates, if we allow some distortion at the output.
However, the MP-AMP problem does
not directly fall into the RD framework, because the quantization error in the current iteration feeds into estimation errors in future iterations. We quantify the interaction between these two forms of error by studying the excess mean squared error (EMSE)
of MP-AMP above the MMSE (EMSE=MSE-MMSE, where MSE denotes the mean squared error).
Our first contribution (Section~\ref{sec:DP}) is to use dynamic programming (DP, cf. Bertsekas~\cite{bertsekas1995}) to find a sequence of coding rates that yields a desired EMSE while achieving the smallest combined cost of
communication and computation; our DP-based scheme is proved to yield optimal coding rates.

Our second contribution (Section~\ref{sec:linRateTh}) is to pose the task of finding the optimal coding rate at each iteration in the low EMSE limit as a convex optimization problem. We prove that the optimal coding rate grows approximately linearly in the low EMSE limit. At the same time, we also provide
the theoretic asymptotic growth rate of the optimal coding rates in the limit of low EMSE. 
This provides practitioners with a heuristic to find a near-optimal coding rate sequence without solving the optimization problem.
The linearity of the  optimal coding rate sequence (defined in Section~\ref{sec:DP}) is also illustrated numerically.
With the rate being approximately linear, we obtain that the combined cost of computation and communication scales as $O(\log^2(1/\text{EMSE}))$.

In Section~\ref{sec:Pareto}, we further consider a rich design space that includes various costs, such as the number of iterations $T$, aggregate coding rate $R_{agg}$, which is the sum of the coding rates in all iterations and is formally defined in~\eqref{eq:R_agg}, and the MSE achieved by the estimation algorithm. In such a rich design space, reducing any cost is likely to incur an increase in other costs, and it is impossible to simultaneously minimize all the costs.
Han et al.~\cite{Han2014} reduce the communication costs, and Ma et al.~\cite{MaBaronNeedell2014} develop an algorithm with reduced computation; both works~\cite{Han2014,MaBaronNeedell2014} achieve a reasonable MSE. However, the optimal trade-offs in this rich design space have not been studied.
Our third contribution is to pose the problem of finding the best trade-offs among the individual costs $T,\ R_{agg}$, and $\text{MSE}$ as a multi-objective optimization problem (MOP), and study the properties of Pareto optimal tuples~\cite{DasDennisPareto1998} of this MOP. These properties are verified numerically using
the DP-based scheme developed in this paper.

Finally, we emphasize that although this paper is presented for the specific framework of MP-AMP,
similar methods could be applied to other iterative distributed  algorithms, such as
consensus averaging~\cite{Frasca2008,Thanou2013}, to obtain the optimal coding rate as well as
optimal trade-offs between communication and computation costs.

The rest of the paper is organized as follows. Section~\ref{sec:setting} provides
background content. Section~\ref{sec:DP} formulates a DP scheme that finds an optimal coding rate. Section~\ref{sec:linRateTh} proves that any optimal coding rate in the low EMSE limit grows approximately linearly as iterations proceed. Section~\ref{sec:Pareto} studies the optimal trade-offs among the computation cost, communication cost, and the MSE of the estimate. Section~\ref{sec:realworld} uses some real-world examples to showcase the different trade-offs between communication and computation costs, and Section~\ref{sec:conclude} concludes the paper.

\section{Background}\label{sec:setting}
\subsection{Centralized linear system using AMP}\label{sec:centralAMP}
In our linear system~\eqref{eq:matrix_channel}, we consider an independent and identically distributed (i.i.d.) Gaussian measurement matrix $\A$, i.e.,
$A_{i,j}\sim\mathcal{N}(0,\frac{1}{M})$, where $\mathcal{N}(\mu,\sigma^2)$
denotes a Gaussian distribution with mean $\mu$ and variance $\sigma^2$.
The signal entries follow an i.i.d. distribution, $f_X(x)$.
The noise entries obey $z_i\sim\mathcal{N}(0,\sigma_Z^2)$, where $\sigma_Z^2$ is the noise variance.

Starting from ${\bf x}_0={\bf 0}$, the AMP framework~\cite{DMM2009} proceeds iteratively according to
\begin{align}
{\bf x}_{t+1}&=\eta_t({\bf A}^{\top}{\bf r}_t+{\bf x}_t)\label{eq:AMPiter1},\\
{\bf r}_t&={\bf y}-{\bf Ax}_t+\frac{1}{\kappa}{\bf r}_{t-1}
\langle \eta_{t-1}'({\bf A}^{\top}{\bf r}_{t-1}+{\bf x}_{t-1})\rangle\label{eq:AMPiter2},
\end{align}
where $\eta_t(\cdot)$ is a denoising function, $\eta_{t}'(\cdot)=\frac{d \eta_t({\cdot})}{d\{\cdot\}}$ is the derivative of $\eta_t(\cdot)$, and~$\langle{\bf u}\rangle=\frac{1}{N}\sum_{i=1}^N u_i$
for any vector~${\bf u}\in\mathbb{R}^N$. The subscript $t$ represents the iteration index, ${\{\cdot\}}^\top$ denotes the matrix transpose operation, and $\kappa=\frac{M}{N}$ is the measurement rate.
Owing to the decoupling effect~\cite{Tanaka2002,GuoVerdu2005,GuoWang2008}, in each AMP iteration~\cite{Bayati2011,Montanari2012,Rush_ISIT2016_arxiv},
the vector~$\f_t={\bf A}^{\top}{\bf r}_t+{\bf x}_t$
in (\ref{eq:AMPiter1}) is statistically equivalent to
the input signal ${\bf x}$ corrupted by AWGN $\w_t$ generated by a source $W\sim \mathcal{N}(0,\sigma_t^2)$,
\begin{equation}\label{eq:equivalent_scalar_channel}
\f_t=\x+\w_t.
\end{equation}
We call~\eqref{eq:equivalent_scalar_channel} the {\em equivalent scalar channel}.
In large systems ($N\rightarrow\infty, \frac{M}{N}\rightarrow \kappa$),\footnote{Note that the results of this paper only hold for large systems.} a useful property of AMP~\cite{Bayati2011,Montanari2012,Rush_ISIT2016_arxiv} is that
the noise variance $\sigma_t^2$ evolves following state evolution (SE):
\begin{equation}
\sigma_{t+1}^2=\sigma^2_Z+\frac{1}{\kappa}\text{MSE}(\eta_t,\sigma_t^2),\label{eq:ori_SE}
\end{equation}
where
$\text{MSE}(\eta_t,\sigma_t^2)=\mathbb{E}_{X,W}\left[\left( \eta_t\left( X+W \right)-X \right)^2\right]$, $\mathbb{E}_{X,W}(\cdot)$ is expectation with respect to (w.r.t.) $X$ and $W$,
and $X$ is the source that generates $\x$. Note that $\sigma_1^2=\sigma_Z^2+\frac{\mathbb{E}[X^2]}{\kappa}$, because of the all-zero initial estimate for $\x$.
Formal statements for SE appear
in prior work~\cite{Bayati2011,Montanari2012,Rush_ISIT2016_arxiv}.

In this paper, we confine ourselves to the Bayesian setting, in which we assume knowledge
of the true prior, $f_X(x)$, for the signal $\x$.
Therefore, throughout this paper we use conditional expectation, $\eta_t(\cdot)=\mathbb{E}[\x|\f_t]$, as the MMSE-achieving
denoiser.\footnote{Tan et al.~\cite{Tan2014} showed that AMP with MMSE-achieving denoisers can be used as a building block for algorithms that minimize arbitrary user-defined error metrics.} The derivative of $\eta_t(\cdot)$, which is continuous, can be easily obtained, and is omitted for brevity.
Other denoisers such as soft thresholding~\cite{DMM2009,Montanari2012,Bayati2011} yield MSEs that are larger than that of the MMSE denoiser, $\eta_t(\cdot)=\mathbb{E}[\x|\f_t]$.
When the true prior for $\x$ is unavailable, parameter estimation techniques
can be used~\cite{MaZhuBaron2016TSP}; Ma et al.~\cite{MaBaronBeirami2015ISIT} study the behavior of AMP when the denoiser uses a mismatched prior.

\subsection{MP-LS using lossy MP-AMP}\label{sec:MP-CS_for_MP-AMP}
In the sensing problem formulated in~\eqref{eq:one-node-meas}, the measurement matrix is stored in a distributed manner in each processor node. Lossy MP-AMP~\cite{HanZhuNiuBaron2016ICASSP} iteratively solves MP-LS using lossily compressed messages:
\begin{equation}
\mbox{Processor nodes:}\ {\bf r}_t^p={\bf y}^p-\A^p\x_t+\frac{1}{\kappa}{\bf r}_{t-1}^p
\omega_{t-1},\label{eq:slave1}
\end{equation}
\begin{equation}
\quad \quad \quad {\bf f}_t^p=\frac{1}{P}\x_t+(\A^p)^{\top}{\bf r}_t^p,\label{eq:slave2}
\end{equation}
\begin{equation}
\mbox{Fusion center:}\ {\bf f}_{Q,t}=\sum_{p=1}^P Q({\bf f}_{t}^p),\ \omega_{t}=\langle d\eta_{t}({\bf f}_{Q,t})\rangle,\label{eq:master0}
\end{equation}
\begin{equation}
 \x_{t+1}=\eta_{t}( {\bf f}_{Q,t}),\label{eq:master}
\end{equation}
where $Q(\cdot)$ denotes quantization, and
an MP-AMP iteration refers to the process from~\eqref{eq:slave1} to~\eqref{eq:master}.
The processor nodes send quantized (lossily compressed) messages, $Q(\f_t^p)$, to the fusion center. The reader might notice that the fusion center also needs to transmit the denoised signal vector $\x_t$ and a scalar $\omega_{t-1}$ to the processor nodes. The transmission of
$\omega_{t-1}$ is negligible, and the fusion center may broadcast $\x_t$ so that naive compression of $\x_t$, such as compression with a fixed quantizer, is sufficient. Hence, we will not discuss possible  compression of messages transmitted by the fusion center.

Assume that we quantize $\f_t^p, \forall p$, and use $C$ bits to encode the quantized vector $Q(\f_t^p)\in\mathbb{R}^N$. The {\em coding rate} is $R=\frac{C}{N}$. We incur an {\em expected distortion}
\begin{equation*}
D_t^p=\mathbb{E}\left[\frac{1}{N}\sum_{i=1}^N(Q(f_{t,i}^p)-f_{t,i}^p)^2\right]
\end{equation*}
at iteration $t$ in each processor node,\footnote{Because we assume that
$\A$ and $\z$ are both i.i.d., the expected distortions are the same over
all $P$ nodes, and can be denoted by $D_t$ for simplicity.
Note also that $D_t=\mathbb{E}[(Q(f_{t,i}^p)-f_{t,i}^p)^2]$
due to $\x$ being i.i.d.}
where $Q(f_{t,i}^p)$ and $f_{t,i}^p$ are the $i$-th entries of the vectors $Q(\f_t^p)$ and $\f_t^p$, respectively,
and the expectation is over $\f_t^p$.
When the size of the problem grows, i.e., $N\rightarrow\infty$, the rate-distortion (RD) function, denoted by $R(D)$, offers the fundamental information theoretic limit on the coding rate $R$ for communicating a long sequence up to distortion $D$~\cite{Cover06,Berger71,GershoGray1993,WeidmannVetterli2012}.
A pivotal conclusion from RD theory is that coding rates can be greatly reduced even if $D$ is small.
The function $R(D)$ can be computed in various ways~\cite{Arimoto72,Blahut72,Rose94}, and can be achieved by an RD-optimal quantization scheme in the limit of large $N$.
Other quantization schemes may require larger coding rates to achieve the same expected distortion $D$.

The goal of this paper is to understand the fundamental trade-offs for MP-LS using MP-AMP. Hence,
unless otherwise stated, we assume that
appropriate vector quantization (VQ)
schemes~\cite{LBG1980,Gray1984,GershoGray1993}, which achieve $R(D)$,
are applied within each MP-AMP iteration, although our analysis is readily extended to practical quantizers such as entropy coded scalar quantization (ECSQ)~\cite{GershoGray1993,Cover06}. (Note that the cost of running quantizers
in each processor node is not considered, because
the cost of processing a bit is usually much smaller than the cost of transmitting it.)
Therefore, the signal {\em at the fusion center} before denoising can be modeled as
\begin{align}
\f_{Q,t}=\sum_{p=1}^P Q(\f_t^p)=\x+\w_t+\n_t,\label{eq:indpt_noises}
\end{align}
where $\w_t$ is the equivalent scalar channel noise~\eqref{eq:equivalent_scalar_channel} and $\n_t$ is the overall quantization error whose entries follow $\mathcal{N}(0,PD_t)$.
Because the quantization error, $\n_t$, is a sum of quantization errors in the $P$ processor nodes,
$\n_t$ resembles Gaussian noise due to the central limit theorem.
Han et al. suggest that SE for lossy MP-AMP~\cite{HanZhuNiuBaron2016ICASSP} (called lossy SE) follows
\begin{equation}\label{eq:SE_Q}
\sigma_{t+1}^2=\sigma^2_Z+\frac{1}{\kappa}\text{MSE}(\eta_t,\sigma_t^2+PD_t),
\end{equation}
where $\sigma_t^2$ can be estimated by
$\widehat{\sigma}_t^2 = \frac{1}{M}\|{\bf r}_t\|_2^2$ with $\|\cdot\|_p$ denoting the $\ell_p$ norm~\cite{Bayati2011,Montanari2012}, and $\sigma_{t+1}^2$ is the variance of $\w_{t+1}$.

The rigorous justification of~\eqref{eq:SE_Q} by extending the framework put forth by Bayati and Montanari~\cite{Bayati2011} and Rush and Venkataramanan~\cite{Rush_ISIT2016_arxiv} is left for future work. Instead, we argue that lossy SE~\eqref{eq:SE_Q} asymptotically tracks the evolution of $\sigma_t^2$ in lossy MP-AMP in the limit of $\frac{PD_t}{\sigma_t^2}\rightarrow 0$.
Our argument is comprised of three parts: ({\em i}) $\w_t$ and $\n_t$~\eqref{eq:indpt_noises} are approximately independent in the limit of $\frac{PD_t}{\sigma_t^2}\rightarrow 0$,   ({\em ii})  $\w_t+\n_t$ is approximately independent of $\x$ in the limit of $\frac{PD_t}{\sigma_t^2}\rightarrow 0$, and ({\em iii}) lossy SE~\eqref{eq:SE_Q} holds if ({\em i}) and ({\em ii}) hold.
The first part ($\w_t$ and $\n_t$ are independent) ensures that we can track the variance of $\w_t+\n_t$ with $\sigma_t^2+PD_t$. The second part ($\w_t+\n_t$ is independent of $\x$) ensures that lossy MP-AMP
follows lossy SE~\eqref{eq:SE_Q} as it falls under the general framework discussed in Bayati and Montanari~\cite{Bayati2011} and Rush and Venkataramanan~\cite{Rush_ISIT2016_arxiv}. Hence, the third part of our argument holds. 
\iftoggle{TSP}{The numerical justification of these three parts appears in an extended version of this paper~\cite{ZhuBaronMPAMP2016ArXiv}.}{The first two parts are backed up by extensive numerical evidence in Appendix~\ref{app:verifyIndpt}, where ECSQ~\cite{GershoGray1993,Cover06} is used; ECSQ
approaches $R(D)$ within 0.255 bits in the high rate limit (corresponds to small distortion)~\cite{GershoGray1993}. Furthermore, Appendix~\ref{app:verifyLossySE} provides extensive numerical evidence to show that lossy SE~\eqref{eq:SE_Q} indeed tracks the evolution of the MSE when $\w_t$ and $\n_t$ are independent and $\w_t+\n_t$ and $\x$ are independent.}

Although lossy SE~\eqref{eq:SE_Q} requires $\frac{PD_t}{\sigma_t^2}\rightarrow 0$, if scalar quantization is used in
a practical implementation,
then lossy SE approximately holds when $\gamma<\frac{2\sigma_t}{\sqrt{P}}$, where $\gamma$ is the quantization bin size of the scalar quantizer (details in 
\iftoggle{TSP}{the extended version of this paper~\cite{ZhuBaronMPAMP2016ArXiv}} {Appendices~\ref{app:verifyIndpt} and~\ref{app:verifyLossySE}}). 
Note that the condition $\gamma<\frac{2\sigma_t}{\sqrt{P}}$ is motivated by Widrow and Koll{\'a}r~\cite{widrow2008quantization}. If appropriate VQ schemes~\cite{LBG1980,Gray1984,GershoGray1993} are used, then we might need milder requirements than $\frac{PD_t}{\sigma_t^2}\rightarrow 0$ in the scalar quantizer case, in order for $\w_t$ and $\n_t$ to be independent and for $\w_t+\n_t$ and $\x$ to be independent.

Denote the coding rate used to transmit $Q(\f^p_t)$ at iteration $t$ by $R_t$. The sequence $\mathbf{R}=(R_1,...,R_T)$ is called
the {\em coding rate sequence}, where $T$ is the total number of MP-AMP iterations. Given
$\mathbf{R}$, the distortion $D_t$ can be evaluated with $R(D)$, and the scalar channel noise variance $\sigma_t^2$ can be evaluated with~\eqref{eq:SE_Q}.
Hence, the MSE for $\mathbf{R}$ can be predicted. The MSE at the last iteration is called the {\em final MSE}.

\section{Optimal Rates Using Dynamic Programming}\label{sec:DP}

In this section, we first define the cost of running MP-AMP. We then use DP to find an optimal coding rate sequence with minimum cost, while achieving a desired EMSE.

\begin{myDef}[Combined cost]\label{def:costFunc}
Define the cost of estimating a signal in an MP system as
\begin{equation}\label{eq:cost}
C^b(\mathbf{R})=b \|\mathbf{R}\|_0+ \|\mathbf{R}\|_1,
\end{equation}
where $\|\mathbf{R}\|_0=T$ is the number of iterations to run, and
$\|\mathbf{R}\|_1$ is
 the aggregate coding rate, denoted also by $R_{agg}$,
\begin{equation}\label{eq:R_agg}
R_{agg}=\| \mathbf{R} \|_1=\sum_{t=1}^T R_t.
\end{equation}
The parameter $b$ is the cost of computation
in one MP-AMP iteration normalized by the cost of transmitting $Q(\f_t^p)$~\eqref{eq:master0} at a coding rate of 1 bit/entry.
Also, the cost at iteration $t$ is
\begin{equation}\label{eq:cost_one_iter}
C^b_t(R_t)  = b\times {\mathbbm{1}}_{R_t \neq 0} + R_t,
\end{equation}
where the indicator function ${\mathbbm{1}}_{\mathcal{A}}$
is 1 if the condition $\mathcal{A}$ is met,
else 0.
Hence, $C^b(\mathbf{R}) = \sum_{t =1}^T C_t^b (R_t)$.
\end{myDef}

In some applications, we may want to obtain a sufficiently small EMSE at minimum cost~\eqref{eq:cost}, where the physical meaning of the cost varies in different problems (cf. Section~\ref{sec:realworld}).
Denote the EMSE at iteration $t$ by $\epsilon_t$. Hence, the {\em final EMSE} at the output of MP-AMP is $\epsilon_T$.

Let us formally state the problem.
Our goal is to obtain a coding rate sequence $\mathbf{R}$ for MP-AMP iterations, which is the solution of the following optimization problem:
\begin{equation}\label{eq:optimizationSetup}
\text{minimize } C^b(\mathbf{R}) \quad \quad
\text{subject to }  \epsilon_T\leq \Delta.
\end{equation}
We now have a definition for the optimal coding rate sequence.

\begin{myDef}[Optimal coding rate sequence]\label{def:optRate}
An optimal coding rate sequence $\mathbf{R}^*$ is a solution of~\eqref{eq:optimizationSetup}.
\end{myDef}

To compute $\mathbf{R}^*$, we derive a dynamic programming (DP)~\cite{bertsekas1995}
scheme, and then prove that it is optimal.

{\bf Dynamic programming scheme:}
Suppose that MP-AMP is at iteration $t$. Define the smallest cost for the $(T-t)$ remaining iterations to achieve the EMSE constraint, $\epsilon_T\leq \Delta$, as $\Phi_{T-t}(\sigma_t^2)$, which is a function of the  scalar channel noise variance at iteration $t$, $\sigma^2_t$~\eqref{eq:indpt_noises}.
Hence, $\Phi_{T-1}(\sigma_1^2)$ is the cost for solving~\eqref{eq:optimizationSetup}, where $\sigma_1^2=\sigma_Z^2+\frac{1}{\kappa}\mathbb{E}[X^2]$ is due to the all-zero initialization of the signal estimate.

DP uses a base case and recursion steps to find $\Phi_{T-1}(\sigma_1^2)$.
In the base case of DP, $T-t=0$, the cost of running MP-AMP is $C^b_T(R_T)=b\times \mathbbm{1}_{R_{T}\neq 0}+R_{T}$~\eqref{eq:cost_one_iter}.
If $\sigma^2_{T}$ is not too large, then there exist some values for $R_T$ that satisfy $\epsilon_T\leq \Delta$; for these $\sigma^2_{T}$ and $R_T$, we have $\Phi_{0}(\sigma^2_{T})=\min_{R_T} C^b_T(R_T)$.
If $\sigma^2_{T}$ is too large, even lossless transmission of $\f_{T}^p$ during the single remaining MP-AMP iteration~\eqref{eq:SE_Q} does not yield an EMSE that satisfies the constraint, $\epsilon_T\leq \Delta$, and
we assign $\Phi_{0}(\sigma^2_{T})=\infty$ for such $\sigma_{T}^2$.

Next, in the recursion steps of DP,
we iterate back in time by decreasing $t$ (equivalently, increasing $T-t$),
\begin{equation}\label{eq:DPrecursion}
 \Phi_{T-t}(\sigma^2_t)\! =\! \min_{\widehat{R}}\left\{ C_t^b(\widehat{R})+ \Phi_{T-(t+1)}(\sigma^2_{t+1}(\widehat{R}))\right\},
\end{equation}
where $\widehat{R}$ is the coding rate used in the current
MP-AMP iteration~$t$,
the equivalent scalar channel noise variance at the fusion center is
$\sigma_t^2$~\eqref{eq:indpt_noises},
and $\sigma^2_{t+1}(\widehat{R})$, which is obtained from (\ref{eq:SE_Q}),
is the variance of the scalar channel noise~\eqref{eq:indpt_noises} in the
next iteration after transmitting $\f_t^p$ at rate $\widehat{R}$.
The terms on the right hand side are the current cost of MP-AMP~\eqref{eq:cost_one_iter}
(including computational and communication costs) and the minimum combined cost
in all later iterations, $t+1,...,T$.

The coding rates $\widehat{R}$ that yield the smallest cost $\Phi_{T-t}(\sigma_t^2)$ for different $t$ and $\sigma^2_t$ are stored in a table $\mathcal{R}(t,\sigma^2_t)$.
After DP finishes, we obtain the coding rate for the first MP-AMP iteration as
$R_1=\mathcal{R}(1,\sigma_Z^2+\frac{1}{\kappa}\mathbb{E}[X^2])$.
Using $R_1$, we calculate $\sigma_t^2$ from~\eqref{eq:SE_Q} for $t=2$ {and
find $R_2=\mathcal{R}(2,\sigma^2_2)$. Iterating from $t=1$ to $T$, we obtain $\mathbf{R}=(R_1,\cdots,R_T)$.

\begin{figure}[t]
\begin{center}
\includegraphics[width=8cm]{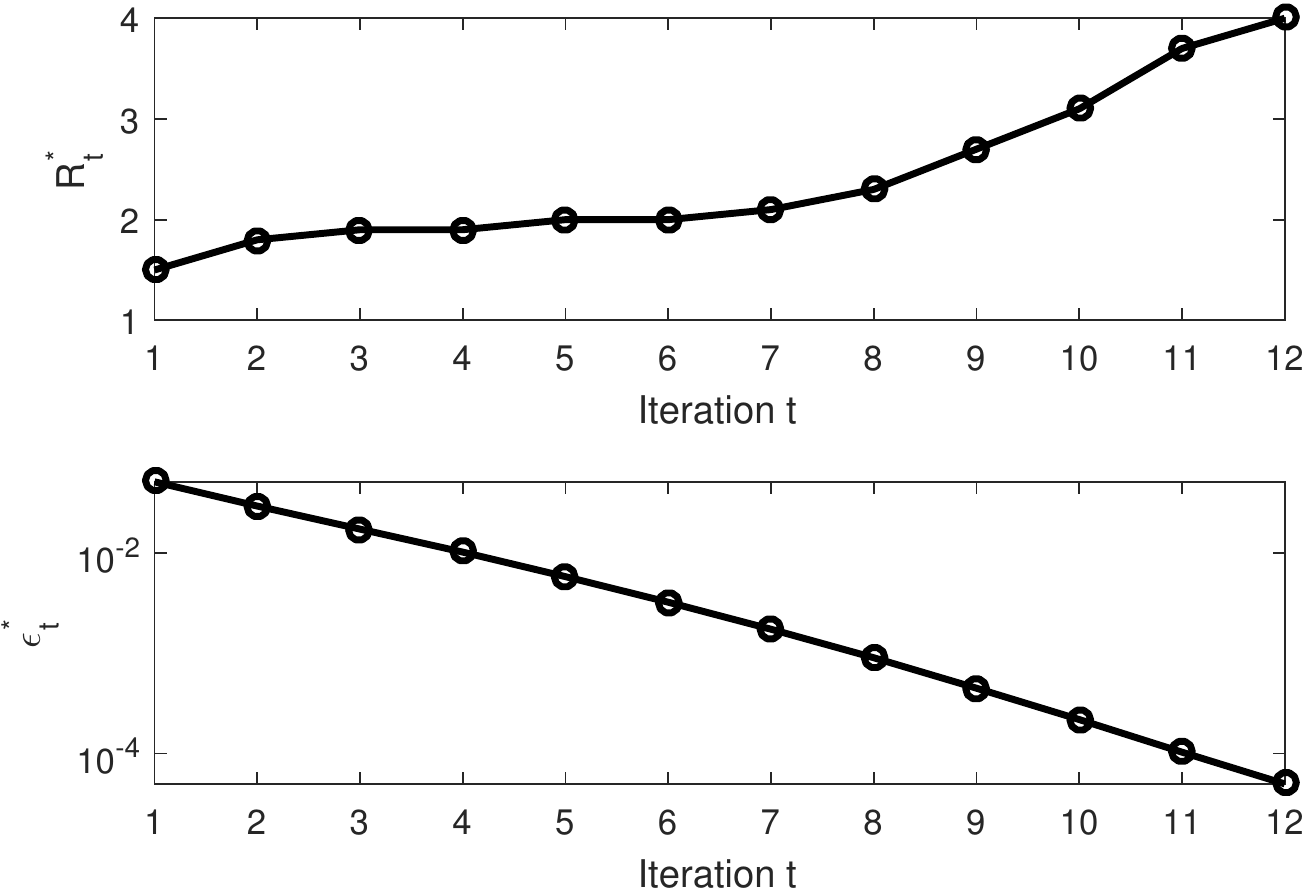}
\end{center}
\caption{The optimal coding rate sequence $\mathbf{R}^*$ (top panel) and optimal EMSE $\epsilon_t^*$ (bottom) given by DP are  shown as functions of $t$. (Bernoulli Gaussian signal~\eqref{eq:BG} with $\rho=0.1$, $\kappa=0.4$, $P=100$, $\sigma_Z^2=\frac{1}{400}$, and $b=2$.)}
\label{fig:RtAndEMSE}
\end{figure}

To be computationally tractable, the proposed DP scheme should operate in
discretized search spaces for $\sigma^2_{\{\cdot\}}$ and $R_{\{\cdot\}}$.
Details about the resolutions of $\sigma^2_{\{\cdot\}}$ and $R_{\{\cdot\}}$ appear in
\iftoggle{TSP}{the extended version  of this paper~\cite{ZhuBaronMPAMP2016ArXiv}}
{Appendix~\ref{app:Integrity}}.

In the following, we state that our DP scheme yields the optimal solution. The proof appears in Appendix~\ref{app:proofDPoptimal}.

\begin{myLemma}\label{lemma:DPoptimal}
The dynamic programming formulation in~\eqref{eq:DPrecursion} yields an optimal coding rate sequence $\mathbf{R}^*$, which is a solution of~\eqref{eq:optimizationSetup} for the discretized search spaces
of $R_t$ and $\sigma_t^2,\ \forall t$.
\end{myLemma}

Lemma~\ref{lemma:DPoptimal} focuses on the optimality of our DP scheme in
discretized search spaces for $R_t$ and $\sigma_t^2$. It can be shown that we can achieve a desired accuracy level in $\mathbf{R}^*$ by adjusting the resolutions of the discretized search spaces for $R_t$ and $\sigma_t^2$.
Suppose that the discretized search spaces for $\sigma^2_{\{\cdot\}}$ and $R_{\{\cdot\}}$ have
$K_1$ and $K_2$ different values, respectively. Then, the computational complexity of our DP scheme is $O(TK_1K_2)$.

{\bf Optimal coding rate sequence given by DP:}
Consider estimating a {\em Bernoulli Gaussian} signal,
\begin{equation}
X=X_BX_G,\label{eq:BG}
\end{equation}
where $X_B\sim \text{Ber}(\rho)$ is a Bernoulli random variable,
$\rho$ is called the {\em sparsity rate} of the signal,
and $X_G\sim {\cal N}(0,1)$;
here we use $\rho=0.1$. Note that the results in this paper apply to priors, $f_X(x)$,
other than~\eqref{eq:BG}.

We run our DP scheme on a problem with relatively small
desired EMSE,  $\Delta=5\times10^{-5}$, in the last iteration $T$.
The signal is measured in an MP platform with $P=100$ processor nodes according to~\eqref{eq:one-node-meas}.
The measurement rate is $\kappa=\frac{M}{N}=0.4$, and the noise variance is $\sigma_Z^2=\frac{1}{400}$. The parameter $b=2$~\eqref{eq:cost}.
We use ECSQ~\cite{GershoGray1993,Cover06} as the quantizer in each processor node, and use  the corresponding relation between the rate $R_t$ and distortion $D_t$ of ECSQ in our DP scheme. Note that we require the quantization bin size to be smaller than $\frac{2\sigma_t}{\sqrt{P}}$, according to Section~\ref{sec:MP-CS_for_MP-AMP}.
Fig.~\ref{fig:RtAndEMSE} illustrates the optimal coding rate sequence $\mathbf{R}^*$ and optimal EMSE $\epsilon_t^*$ given by DP as functions of the iteration number $t$.

It is readily seen that after the first 5--6 iterations the coding rate seems near-linear.
The next section proves that any optimal coding rate sequence $\mathbf{R}^*$
is approximately linear in the limit of EMSE$\rightarrow 0$.
However, our proof involves the large $t$ limit, and does not provide insights for small $t$.
We ran DP for various configurations.
Examining all $\mathbf{R}^*$ from our DP results, we notice that the coding rate is
monotone non-decreasing, i.e., $R^*_1\leq R^*_2\leq \cdots\leq R^*_T$. This seems intuitive, because in
early iterations of (MP-)AMP, the scalar channel noise $\w_t$ is large, which does not require
transmitting $\f_t^p$ (cf.~\eqref{eq:slave2}) at high fidelity. Hence, a
low rate $R^*_t$ suffices. As the iterations proceed, the scalar channel noise
$\w_t$ in~\eqref{eq:indpt_noises} decreases, and the large quantization
error $\n_t$ would be unfavorable for the final MSE.
Hence, higher rates are needed in later iterations.

\section{Properties of Optimal Coding Rate Sequences}\label{sec:linRateTh}
\subsection{Intuition}\label{sec:intuition}

We start this section by providing some brief intuitions about why optimal coding rate sequences are approximately linear when the EMSE is small.

Consider a case where we aim to reach a low $\text{EMSE}$. Montanari~\cite{Montanari2012}
provided a geometric interpretation of the relation between the MSE performance of AMP at
iteration $t$ and
the denoiser $\eta_t(\cdot)$ being used.\footnote{We will also provide such an interpretation in Section~\ref{sec:geoInterp}.}
In the limit of small $\text{EMSE}$, the $\text{EMSE}$ decreases by a nearly-constant multiplicative factor
per AMP iteration, yielding a geometric decay of the EMSE.
In MP-AMP, in addition to the equivalent scalar channel noise $\w_t$, we have additive quantization error $\n_t$~\eqref{eq:indpt_noises}.
In order for the $\text{EMSE}$ in an MP-AMP system to decay geometrically, the distortion $D_t$ must decay at least as quickly. To obtain this geometric decay in $D_t$,
recall that in the high rate limit, the distortion-rate function typically takes the form $D(R)\approx C_1 2^{-2R}$~\cite{GrayNeuhoff1998} for some positive constant $C_1$.
We propose for $R_t$ to have the form,
$R_t\approx C_2+C_3 t$,
where $C_2$ and $C_3$ are constants. In the remainder of this section, we first discuss the geometric interpretation of AMP state evolution, followed by our results about the linearity of optimal coding rate sequences. The detailed proofs appear in the appendices.

\begin{figure*}[t]
\centering
  \subfigure[]{
    \label{fig:losslessSE} 
    \includegraphics[height=0.23\textwidth]{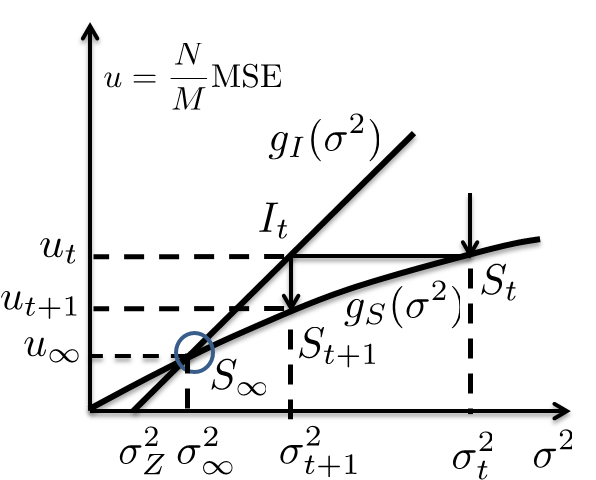}}
    \subfigure[]{
    \label{fig:zoomIn} 
    \includegraphics[height=0.23\textwidth]{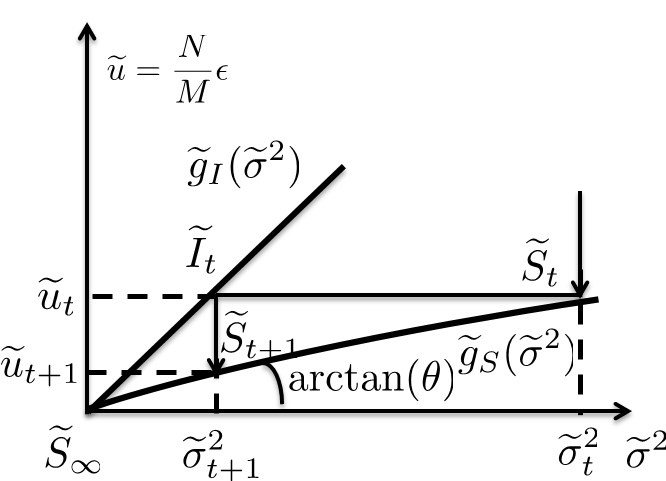}}
  \subfigure[]{
    \label{fig:lossySE} 
    \includegraphics[height=0.23\textwidth]{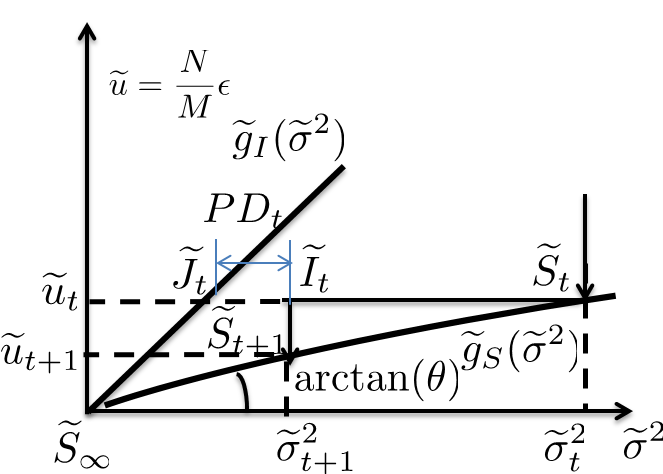}}
\caption{Geometric interpretation of SE. In all panels, the thick solid curves correspond to $g_I(\cdot)$ and $g_S(\cdot)$, and their offset versions $\widetilde{g}_I(\cdot)$ and $\widetilde{g}_S(\cdot)$. The solid lines with arrows correspond to the SE of AMP.   Dashed lines without arrows are auxiliary lines. Panel (a): Illustration of centralized SE. Panel (b): Zooming in to the small region just above point $S_\infty$. Panel (c): Illustration of lossy SE.}\label{fig:AMP_SE_geo}
\end{figure*}

\subsection{Geometric interpretation of AMP state evolution}\label{sec:geoInterp}
{\bf Centralized SE:}
The equivalent scalar channel of AMP is given by~\eqref{eq:equivalent_scalar_channel}.
We re-write the centralized AMP SE~\eqref{eq:ori_SE} as follows~\cite{DMM2009,Bayati2011,Rush_ISIT2016_arxiv},
\begin{equation}\label{eq:SE}
\underbrace{\sigma_{t+1}^2-\sigma_Z^2}_{g_I(\sigma_{t+1}^2)}=\underbrace{\frac{N}{M}\text{MSE}_{\eta_t}(\sigma_t^2)}_{g_S(\sigma_t^2)},
\end{equation}
where $\text{MSE}_{\eta_t}(\sigma_t^2)$ denotes the MSE after denoising $\f_t$~\eqref{eq:equivalent_scalar_channel}
using $\eta_t(\cdot)$.
The functions $g_I(\cdot)$ and $g_S(\cdot)$ are illustrated in Fig.~\ref{fig:losslessSE}
with solid curves; the meanings of $I$ and $S$ will become clear below.
We see that $g_I(\sigma_t^2)$ is an affine function with unit slope, whereas $g_S(\sigma_t^2)$ is generally a nonlinear function of $\sigma_t^2$ (see Fig.~\ref{fig:losslessSE}). The lines with arrows illustrate the state evolution (SE). Details appear below.

In Fig.~\ref{fig:losslessSE}, we present a
geometric interpretation of SE. The horizontal axis is
the scalar channel noise variance $\sigma^2$ and the vertical axis represents the scaled MSE, $u=\frac{N}{M}\text{MSE}$.
Let $S_t = (\sigma^2_t, u_t)$ be the {\em state} point that is reached by SE in
iteration $t$.
We follow the SE trajectory $S_t \to I_t \to S_{t+1} \to \cdots$
in Fig.~\ref{fig:losslessSE},  
where $I_t = (\sigma^2_{t+1}, u_t)$ represents the {\em intermediate} point in the transition between states $S_t$ and $S_{t+1}$ corresponding to iterations $t$ and $t+1$, respectively. 
Observe that the points $S_t$ and $I_t$ have the same ordinate ($u_t$), while $S_{t+1}$ and $I_t$ have the same abscissa ($\sigma_{t+1}^2$), which are related as
$\sigma^2_{t+1} = g^{-1}_I(u_t)$
and
$u_{t+1} = g_S(\sigma^2_{t+1})$.
As $t$ grows, $\sigma^2_t$  converges to $\sigma^2_\infty$, which is the abscissa of the point $S_\infty$. The ordinate of point $S_\infty$ is $u_\infty=\frac{N}{M}$MSE$_\infty$, where $\text{MSE}_\infty = \text{MMSE}$. If we stop the algorithm at iteration $T$, or equivalently at point $S_T = (\sigma^2_T, u_T)$,
the corresponding MSE, MSE$_T$, has an EMSE of $\epsilon_T=\text{MSE}_T-\text{MMSE}$.

In Fig.~\ref{fig:zoomIn}, we zoom into the neighborhood of point $S_\infty$. To make the presentation more concise, we vertically offset $g_I(\cdot)$ and $g_S(\cdot)$ by $\frac{N}{M}$MMSE and horizontally offset them by $\sigma_\infty^2$; we call the resulting functions $\widetilde{g}_I(\cdot)$ and $\widetilde{g}_S(\cdot)$, respectively. Hence, the vertical axis in Fig.~\ref{fig:zoomIn} represents the scaled EMSE, $\widetilde{u}=\frac{N}{M}\text{EMSE}=\frac{N}{M}\epsilon$, and we have $\widetilde{g}_I(\widetilde{\sigma}^2_t)=g_I(\widetilde{\sigma}^2_t+\sigma_\infty^2)-\frac{N}{M}$MMSE and $\widetilde{g}_S(\widetilde{\sigma}^2_t)=g_S(\widetilde{\sigma}^2_t+\sigma_\infty^2)-\frac{N}{M}$MMSE.
Observe that $\widetilde{g}_I(0) = \widetilde{g}_S(0) = 0$.
Additionally, the slope of $\widetilde{g}_I(\widetilde{\sigma}^2_t)$ is $\widetilde{g}_I'(\widetilde{\sigma}^2_t)=1$, where $\widetilde{g}_I'(\cdot)$ is the first-order derivative of $\widetilde{g}_I(\cdot)$ w.r.t. $\widetilde{\sigma}^2_t$ (Fig.~\ref{fig:zoomIn}).
Because the MSE function for the MMSE-achieving denoiser is continuous and differentiable twice~\cite{WuVerdu2011}, we can invoke Taylor's theorem to express
\begin{equation}
\widetilde{g}_S(\widetilde{\sigma}^2_t)
=\widetilde{g}_S'(0)\widetilde{\sigma}^2_t+\frac{1}{2}\widetilde{g}_S''(\zeta_t)\widetilde{\sigma}^4_t,
\label{eq:Taylor}
\end{equation}
where $\zeta_t\in (0,\widetilde{\sigma}^2_t),$ and $\widetilde{g}_S'(\widetilde{\sigma}^2_t)$ and $\widetilde{g}_S''(\widetilde{\sigma}^2_t)$ are the first- and second-order derivatives of $\widetilde{g}_S(\cdot)$ w.r.t.  $\widetilde{\sigma}^2_t$, respectively.
Due to continuity and differentiability of the denoising function, $\widetilde{g}_S(\cdot)$ is invertible in a neighborhood around $0$, and its inverse is denoted by $\widetilde{g}^{-1}_S(\cdot).$
Invoking Taylor's theorem,
\begin{equation}
\widetilde{g}_S^{-1}(\widetilde{u}_t)=(\widetilde{g}_S^{-1})'(0) \widetilde{u}_t+\frac{1}{2}(\widetilde{g}_S^{-1})''(\zeta_t)\widetilde{u}_t^2,
\label{eq:Taylor_inverse}
\end{equation}
where $\zeta_t\in (0,\widetilde{u}_t)$, and
$(\widetilde{g}^{-1}_S)'(\widetilde{u}_t)$ and
$(\widetilde{g}^{-1}_S)''(\widetilde{u}_t)$ are the first- and second-order derivatives of $\widetilde{g}^{-1}_S(\cdot)$ w.r.t.  $\widetilde{u}_t$, respectively. When $t\rightarrow \infty$, $\widetilde{\sigma}^2_t \to 0$ and $\widetilde{u}_t \to 0$,
and the higher-order terms become
$\frac{1}{2}\widetilde{g}_S''(\xi_t)\widetilde{\sigma}_t^4=O(\widetilde{\sigma}_t^4)$ and
$\frac{1}{2}(\widetilde{g}_S^{-1})''(\zeta_t)\widetilde{u}_t^2=O(\widetilde{u}_t^2)$. 
In other words, both $\widetilde{g}_S(\widetilde{\sigma}^2_t)$ and $\widetilde{g}^{-1}_S(\widetilde{u}_t)$ become approximately linear functions, as shown in Fig.~\ref{fig:zoomIn}. We further denote the slope of $\widetilde{g}_S(0)$ by $\theta$, i.e.,
\begin{equation}
\theta=\widetilde{g}_S'(0) = \frac{1}{(\widetilde{g}_S^{-1})'(0)}.
\label{eq:theta}
\end{equation}

To calculate the slope $\theta$, we first calculate the scalar channel noise variance for point $S_\infty$, $\sigma_\infty^2$, by using replica analysis~\cite{ZhuBaronCISS2013,Krzakala2012probabilistic},\footnote{The outcome of replica analysis~\cite{ZhuBaronCISS2013,Krzakala2012probabilistic} is close to simulating SE~\eqref{eq:SE} with a large number of iterations.} and obtain
$\theta=g_S'(\sigma_\infty^2)=\widetilde{g}_S'(0)$.
Moreover, the slope of $\widetilde{g}_S(0)$ satisfies $\theta=\widetilde{g}_S'(0)\in(0,1)$; otherwise, the curves $\widetilde{g}_I(\cdot)$ and $\widetilde{g}_S(\cdot)$ would not intersect at point $S_\infty$.

{\bf Lossy SE:}
Considering lossy SE~\eqref{eq:SE_Q}, we have
\begin{equation}\label{eq:lossySE}
\underbrace{\sigma_{t+1}^2-\sigma_Z^2}_{g_I(\sigma_{t+1}^2)}=\underbrace{\frac{N}{M}\text{MSE}_{\eta_t}(\sigma_t^2+PD_t)}_{g_S(\sigma_t^2+PD_t)},
\end{equation}
where $P$ is the number of processor  nodes in an MP network, and $D_t$ is the expected
distortion incurred by each node at iteration $t$.
Note that lossy SE has not been rigorously proved in the literature, although we argued in Section~\ref{sec:MP-CS_for_MP-AMP} that it tracks the evolution of the equivalent scalar channel noise variance $\sigma_t^2$ when $D_t\ll \frac{1}{P}\sigma_t^2$.

We notice the additional term $PD_t$, which corresponds to the distortion {\em at the fusion center}.
Because the $P$ nodes transmit their signals $\f_t^p$
with distortion $D_t$, and their messages are independent,
the fusion center's signal has distortion $PD_t$.
The lines with arrows in Fig.~\ref{fig:lossySE} illustrate the lossy SE after vertically offsetting $g_I(\cdot)$ and $g_S(\cdot)$ by $\frac{N}{M}$MMSE and horizontally offsetting $g_I(\cdot)$ and $g_S(\cdot)$ by $\sigma_\infty^2$.
After arriving at point $\widetilde{S}_t$, we move horizontally to
$\widetilde{J}_t$, and obtain the ordinate of $\widetilde{I}_t$, $\widetilde{u}_t$, from $\widetilde{g}_S(\widetilde{\sigma}_t^2+PD_t)=\widetilde{u}_t$. Geometrically,
SE is dragged to the right by distance $PD_t$ from point $\widetilde{J}_t$ to
$\widetilde{I}_t$, and then SE descends from $\widetilde{I}_t$
to $\widetilde{S}_{t+1}$.

\subsection{Asymptotic linearity of the optimal coding rate sequence}\label{sec:linearRateTheorem_subsection}

Recall from~\eqref{eq:Taylor} that
 $\lim_{t\rightarrow\infty}\widetilde{\sigma}^2_t=0$. 
Hence, as $t$ grows, $f_{t,i}$~\eqref{eq:equivalent_scalar_channel} converges in distribution to
$x_i +\mathcal{N}(0, \sigma^2_{\infty})$.
Therefore, the RD function converges to some fixed function as
$t$ grows. For large coding rate $R$, this function} has the form
\begin{equation}\label{eq:DR}
R_t=\frac{1}{2}\log_2\l(\frac{C_1}{D_t}\r) (1+ o_t(1)),
\end{equation}
for some constant $C_1$ that does not depend on $t$~\cite{GrayNeuhoff1998}.
Note that the assumption of $\widetilde{\sigma}^2_t$ being small implicitly requires the
coding rate used in the corresponding iteration to be large.

For an optimal coding rate sequence $\mathbf{R}^*$, we call the distortion $D_t^*$, derived from~\eqref{eq:DR}, incurred by the optimal code rate $R_t^*$ at a certain iteration $t$ the {\em optimal distortion}. Correspondingly,
we call the EMSE achieved by MP-AMP with $\mathbf{R}^*$,
denoted by $\epsilon_t^*$, the {\em optimal EMSE} at iteration $t$.
In the following, we state our main results on the optimal coding rate, the optimal distortion, and the optimal EMSE.

\begin{myTheorem}[Linearity of the coding rate sequence]\label{th:optRateLinear}
Supposing that lossy SE~\eqref{eq:lossySE} holds, we have 
\begin{equation}
\lim_{t \to \infty} \frac{D_{t+1}^*}{D_t^*} = \theta,
\label{eq:theorem1-1}
\end{equation}
where $\theta$ is defined in~\eqref{eq:theta}. Further,
\begin{equation}\label{eq:theorem1}
\lim_{t\rightarrow\infty} \l(R^*_{t+1} - R^*_{t }\r)= \frac{1}{2}\log_2 \l(\frac{1}{\theta}\r).
\end{equation}
\end{myTheorem}

Theorem~\ref{th:optRateLinear} is
proved in Appendix~\ref{app:optRateLinear}.

\begin{myRemark}
Define the additive growth rate of an optimal coding rate sequence $\mathbf{R}^*$ at iteration $t$ as $R_{t+1}^*-R_{t}^*$.
Theorem~\ref{th:optRateLinear} not only shows that any optimal coding rate sequence grows approximately linearly in the low EMSE limit, but also provides a way to calculate its additive growth rate in the low EMSE limit. Hence, if the goal is to achieve a low EMSE, practitioners could simply use a coding rate sequence that has a fixed coding rate in the first few iterations and then increases linearly with additive growth rate $\frac{1}{2}\log_2\l(\frac{1}{\theta}\r)$.
\end{myRemark}

The following theorem provides ({\em i}) the relation between the optimal distortion $D_{t+1}^*$ and the optimal EMSE $\epsilon_t^*$ in the large $t$ limit, and ({\em ii}) the convergence rate of the optimal EMSE $\epsilon_t^*$.
\begin{myTheorem}\label{th:convergence}
Assuming that lossy SE~\eqref{eq:lossySE} holds, we have
\begin{equation}\label{eq:theorem2_2}
\lim_{t\rightarrow\infty} \frac{D_t^*}{\epsilon_{t}^*} =0.
\end{equation}
Furthermore, the convergence rate of the optimal EMSE is
\begin{equation}\label{eq:theorem2_1}
\lim_{t\rightarrow\infty} \frac{\epsilon_{t+1}^*}{\epsilon_{t}^*}=\theta.
\end{equation}
\end{myTheorem}

Theorem~\ref{th:convergence} is proved in Appendix~\ref{app:convergence}. Note that $\lim_{t\rightarrow\infty} \frac{D_t^*}{\epsilon_{t}^*} =0$ meets the
requirement $\frac{PD_t}{\sigma_t^2}\rightarrow 0$ discussed
in Section~\ref{sec:MP-CS_for_MP-AMP}.
Extending Theorems~\ref{th:optRateLinear} and~\ref{th:convergence}, we have the following result.

\setcounter{myCoro}{2}

\begin{myCoro}
Assuming that lossy SE~\eqref{eq:SE_Q} holds, the combined computation and communication cost~\eqref{eq:cost} scales as $O(\log^2(1/\Delta))$, $\forall b>0$, where $\Delta$ is the desired EMSE.
\end{myCoro}
\begin{IEEEproof}
Given Theorem~\ref{th:convergence},
we obtain that the optimal EMSE, $\epsilon_t^*$,
indeed decreases geometrically in the large $t$ limit (as a reminder, we provided such intuition in Section~\ref{sec:intuition}). Considering~\eqref{eq:R_agg} and Theorem~\ref{th:optRateLinear}, the total computation and communication cost~\eqref{eq:cost} for running $T$ iterations is
$C^b(\mathbf{R^*})=O(T^2)=O(\log^2(1/\epsilon_T^*)) = O(\log^2(1/\Delta))$.
\end{IEEEproof}

\begin{myRemark}
The key to the proofs of Theorems~\ref{th:optRateLinear} and~\ref{th:convergence} is lossy SE~\eqref{eq:lossySE}. We expect that the linearity of the optimal coding rate sequence could be extended to other iterative distributed algorithms provided that ({\em i}) they have formulations similar to lossy SE~\eqref{eq:lossySE} that track their estimation errors
and ({\em ii}) their estimation errors converge geometrically.
Moreover, formulations that track the estimation error in such
algorithms might require less restrictive constraints than AMP. For example,
consensus averaging~\cite{Frasca2008,Thanou2013}
only requires i.i.d. entries in the vector that each node in the network averages.
\end{myRemark}

\begin{figure}
  \centering
  \includegraphics[width=8cm]{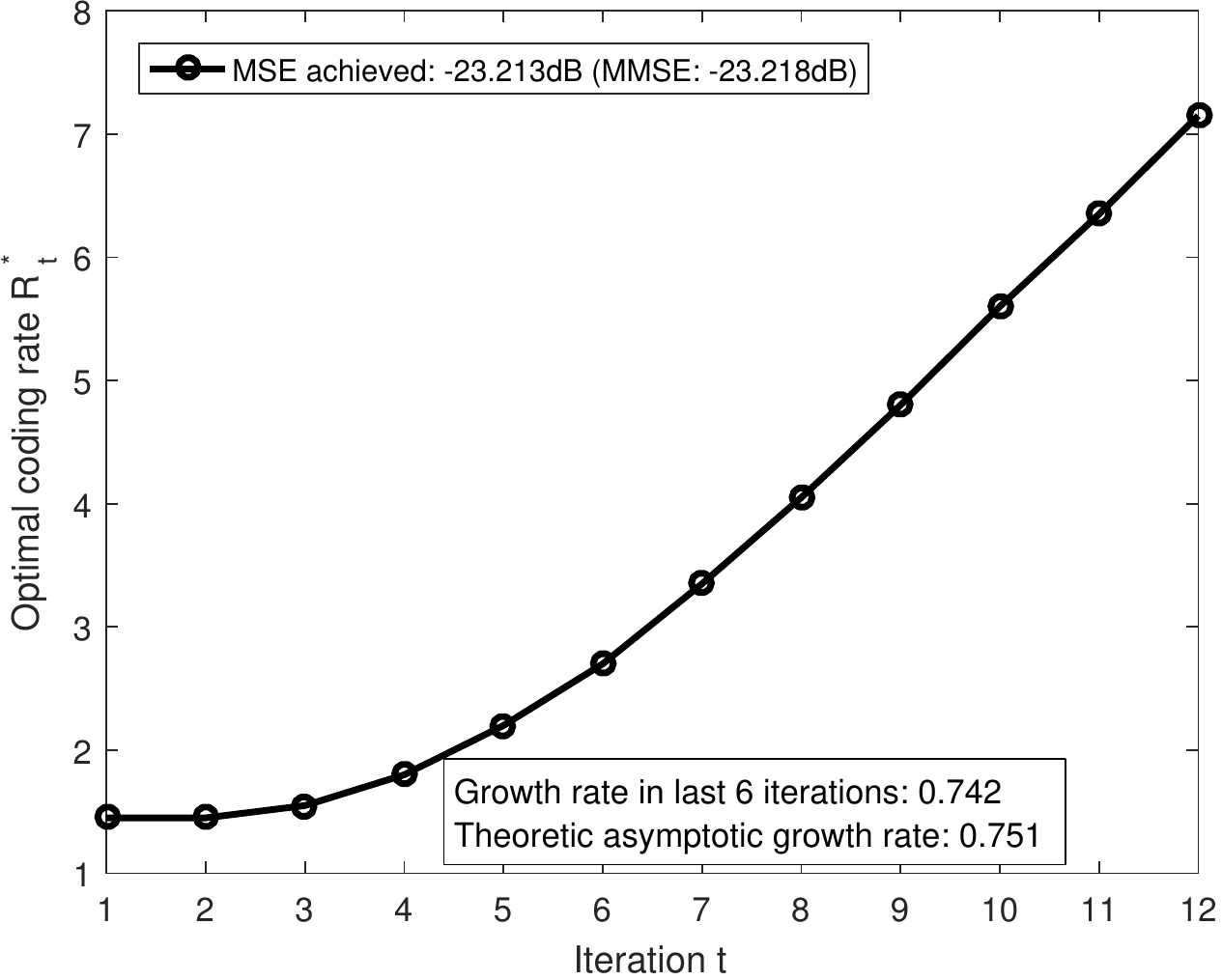}
  \caption{Comparison of the additive growth rate of the optimal coding rate sequence given by DP at low EMSE and the asymptotic  additive growth rate $\frac{1}{2}\log_2\l(\frac{1}{\theta}\r)$. (Bernoulli Gaussian signal~\eqref{eq:BG} with $\rho=0.2,\ \kappa=1,\ P=100, \sigma_Z^2=0.01,\ b=0.782$.)}\label{fig:asympSlope}
\end{figure}

\subsection{Comparison of DP results to Theorem~\ref{th:optRateLinear}}

We run DP (cf. Section~\ref{sec:DP}) to find an optimal coding rate sequence $\mathbf{R}^*$ for the setting of $P=100$ nodes, a Bernoulli Gaussian signal~\eqref{eq:BG} with sparsity rate $\rho=0.2$, measurement rate $\kappa=1$, noise variance $\sigma_Z^2=0.01$, and  parameter $b=0.782$. The goal is to achieve a desired EMSE of 0.005 dB, i.e.,
$10\log_{10}\l(1 + \frac{\Delta}{\text{MMSE}}\r)=0.005$.
We use ECSQ~\cite{GershoGray1993,Cover06} as the quantizer in each processor node and
use the corresponding relation between the
rate $R_t$ and distortion $D_t$ of ECSQ in the DP scheme. Note that we require the
quantization bin size $\gamma$ to be smaller than
$\frac{2\sigma_t}{\sqrt{P}}$, according to Section~\ref{sec:MP-CS_for_MP-AMP}. We know that ECSQ achieves a coding rate
within an additive constant of the RD function $R(D)$~\cite{GershoGray1993}. Therefore, the additive
growth rate of the optimal coding rate sequence
obtained for ECSQ will be the same as the additive growth rate if the RD relation is modeled by $R(D)$~\cite{Cover06,Berger71,GershoGray1993,WeidmannVetterli2012}.

The resulting optimal coding rate sequence is plotted in Fig.~\ref{fig:asympSlope}. The additive growth rate of the last six iterations is $\frac{1}{6}(R_{12}^*-R_{6}^*)=0.742$, and the asymptotic additive growth rate according to Theorem~\ref{th:optRateLinear} is $\frac{1}{2}\log_2\l(\frac{1}{\theta}\r)\approx 0.751$.
Note that we use $\Delta R_t=0.05$ in the discretized search space for $R_t$. Hence, the discrepancy of 0.009 between the additive growth rate from the simulation and the asymptotic additive growth rate is
within our numerical precision.
In conclusion, our numerical result matches the theoretical prediction of Theorem~\ref{th:optRateLinear}.

\section{Achievable Performance Region}\label{sec:Pareto}

Following the discussion of Section~\ref{sec:setting}, we can see that the lossy compression of
${\bf f}_t^p, \forall p \in \{1,...,P\}$, can reduce communication costs. On the other hand,
the greater the savings in the coding rate sequence $\mathbf{R}$, the worse the final MSE is expected to be.
If a certain level of final MSE is desired
despite a small coding rate budget, then more iterations $T$ will be needed.
As mentioned above, there is a trade-off between $T$, $R_{agg}$, and the final
MSE, i.e., $\text{MMSE} + \Delta$, and there is no
solution that minimizes them simultaneously.
To deal with such trade-offs, which implicitly correspond to sweeping
$b$ in~\eqref{eq:cost} in a multi-objective optimization (MOP) problem, it is customary to think about
{\em Pareto optimality}~\cite{DasDennisPareto1998}.

\subsection{Properties of achievable region}\label{sec:property}

For notational convenience, denote
the set of all MSE values achieved by the pair $(T,R_{agg})$
for some parameter $b$~\eqref{eq:cost} by ${\cal E}(T, R_{agg})$.
Within $(T,R_{agg})$, let the smallest MSE be $\text{MSE}^*(T,R_{agg})$.
We now define the achievable set $\cal C$,
$$
{\cal C} := \{(T,R_{agg}, \text{MSE}) \in \mathbb{R}_{\geq 0}^3: \text{MSE} \in {\cal E}(T, R_{agg})\},
$$
where $\mathbb{R}_{\geq 0}$ is the set of non-negative real numbers.
That is, ${\cal C}$ contains all tuples $(T,R_{agg}, \text{MSE})$ for which
some instantiation of MP-AMP estimates the signal at the
desired MSE level using $T$ iterations and aggregate coding rate $R_{agg}$.

\begin{figure*}[t]
\centering
  \subfigure[]{
    \label{fig:rateVSiter} 
    \includegraphics[width=0.32\textwidth]{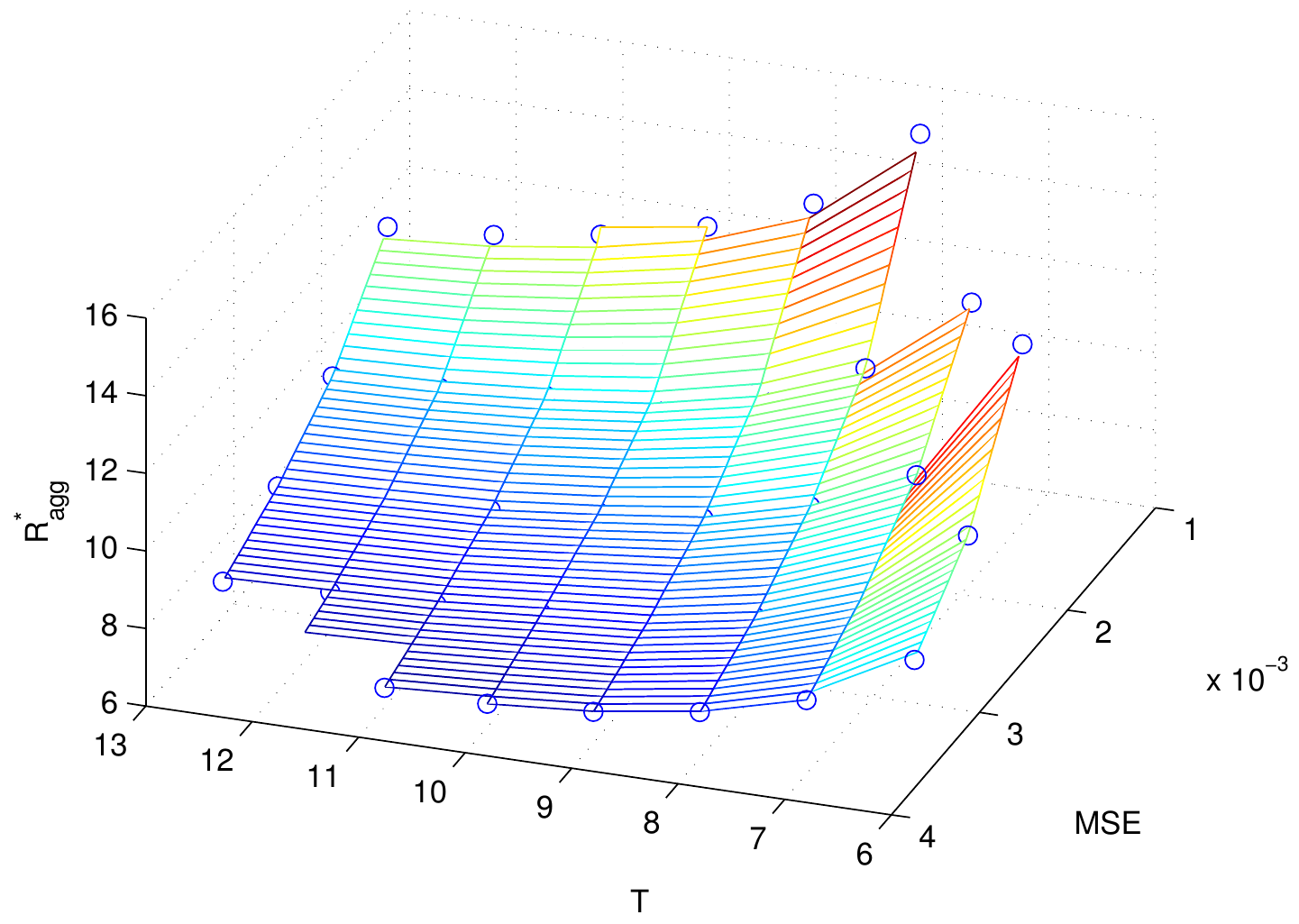}}
  \subfigure[]{
    \label{fig:Pareto2d_fixT} 
    \includegraphics[width=0.3\textwidth]{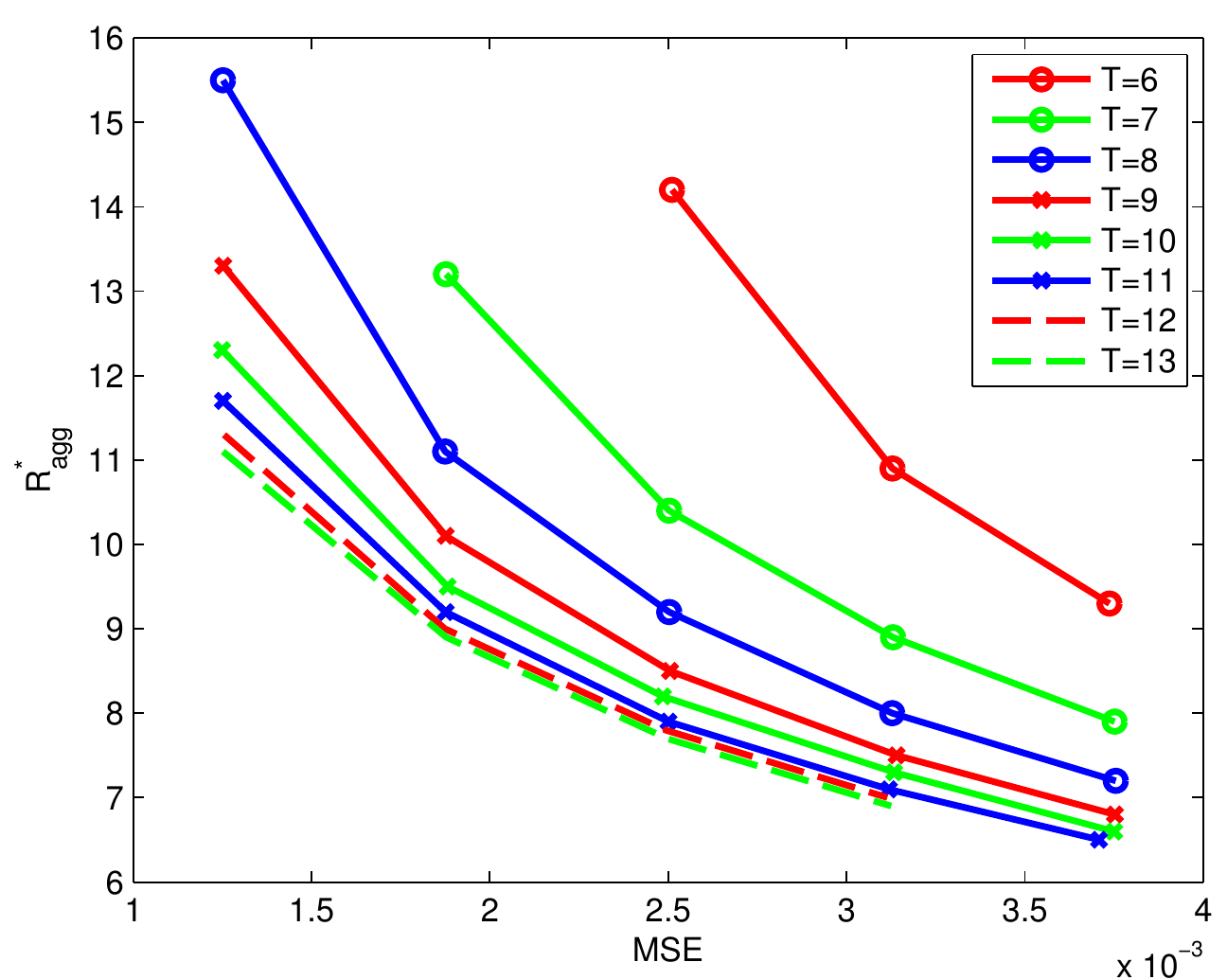}}
  \subfigure[]{
    \label{fig:Pareto2d_fixMSE} 
    \includegraphics[width=0.3\textwidth]{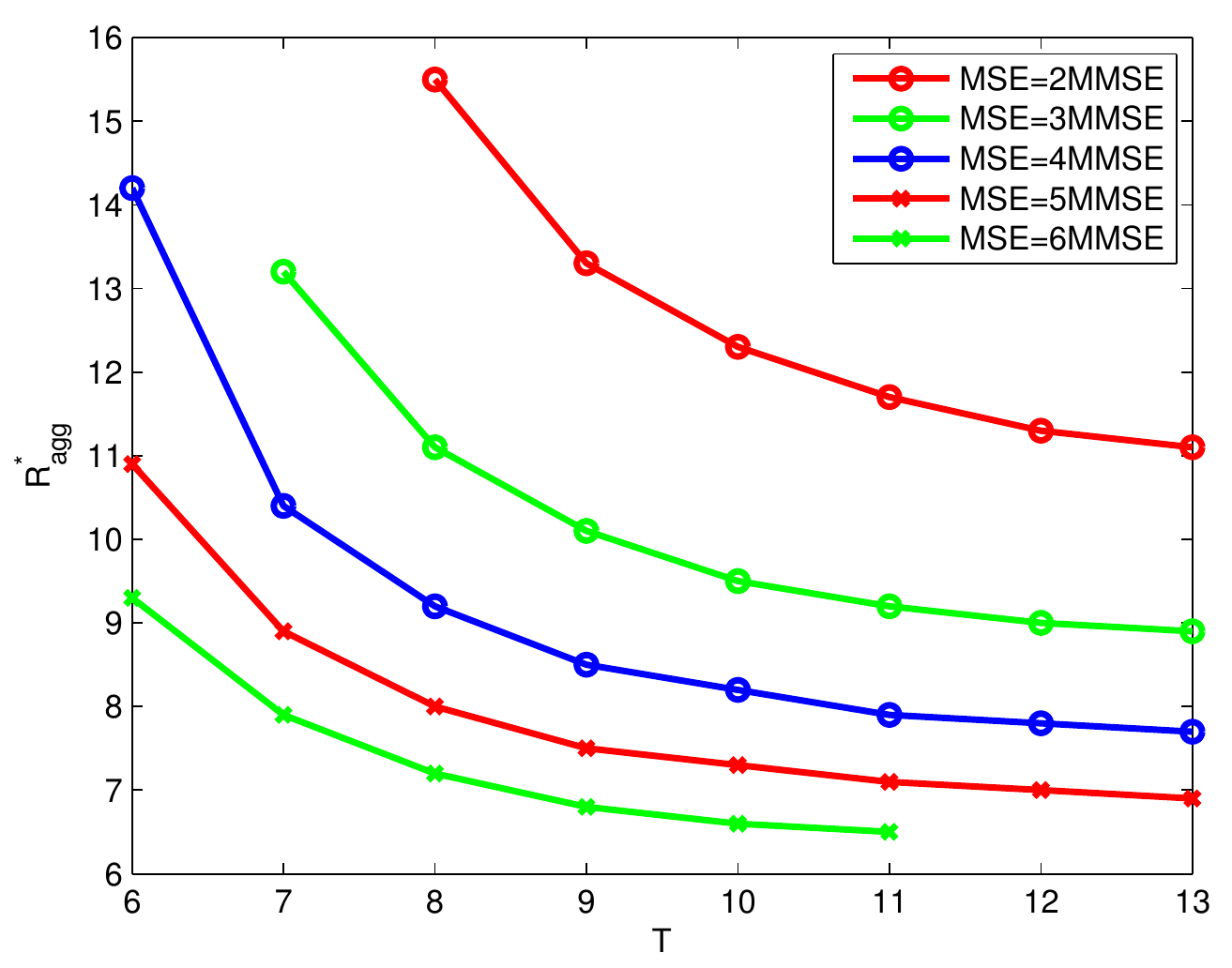}}
\caption{Pareto optimal results provided by DP under a variety of parameters
$b$~\eqref{eq:cost}: (a) Pareto optimal surface, (b) Pareto optimal aggregate coding rate $R_{agg}^*$~\eqref{eq:R_agg} versus the achieved MSE for different optimal MP-AMP iterations $T$, and (c) Pareto optimal $R_{agg}^*$~\eqref{eq:R_agg} versus the number of iterations $T$ for different optimal MSEs.
The signal is Bernoulli Gaussian~\eqref{eq:BG} with $\rho=0.1$. ($\kappa=0.4$, $P=100$, and $\sigma_Z^2=\frac{1}{400}$.)}\label{fig:Pareto}
\end{figure*}

\begin{myDef}\label{def:Pareto}
{\em  The point $\mathcal{X}_1\in\mathcal{C}$ is said to dominate another point $\mathcal{X}_2\in\mathcal{C}$, denoted by $\mathcal{X}_1\prec \mathcal{X}_2$, if $T_1\leq T_2$, $R_{agg_1}\leq R_{agg_2}$, and $\text{MSE}_1\leq \text{MSE}_2$. A point $\mathcal{X}^*\in \mathcal{C}$ is
Pareto optimal if there does not exist $\mathcal{X}\in \mathcal{C}$ satisfying $\mathcal{X}\prec \mathcal{X}^*$.
Furthermore, let $\mathcal{P}$ denote the set of all Pareto optimal points,}
\begin{equation}\label{eq:setP}
\mathcal{P} := \{\mathcal{X}\in \mathcal{C}: \text{$\mathcal{X}$ is Pareto optimal}\}.
\end{equation}
\end{myDef}

In words, the tuple $(T,R_{agg},\text{MSE})$ is Pareto optimal if no other tuple
$(\widehat{T},\widehat{R}_{agg},\widehat{\text{MSE}})$ exists such that
$\widehat{T}\leq T$, $\widehat{R}_{agg}\leq R_{agg}$, and $\widehat{\text{MSE}}\leq \text{MSE}$.
Thus, the Pareto optimal tuples belong to the boundary of $\cal C$.

We extend the definition of the number of iterations $T$ to a probabilistic one. To do so, suppose
that the number of iterations is drawn from a probability distribution $\pi$ over $\mathbb{N}$, such that $\sum_{i=1}^{\infty} \pi_i = 1$. Of course, this definition contains a deterministic $T = j$ as a special case with $\pi_j = 1$ and $\pi_i =0$ for all $i \neq j$.
Armed with this definition of Pareto optimality and the probabilistic definition of the number of iterations, we have the following lemma.

\begin{myLemma}\label{th:convex1}
{\it For a fixed noise variance $\sigma^2_Z$, measurement rate $\kappa$, and $P$ processor nodes in MP-AMP, the achievable set $\mathcal{C}$ is a convex set.}
\end{myLemma}

\begin{IEEEproof}
We need to show that for any $(T^{(1)},$ $R^{(1)}_{agg},\text{MSE}^{(1)})$, $(T^{(2)},R^{(2)}_{agg},\text{MSE}^{(2)})$ $\in \mathcal{C}$ and any $0<\lambda<1$,
\begin{equation}\label{eq:suff1}
\begin{split}
(\lambda T^{(1)} + (1-\lambda)T^{(2)}&,\lambda R^{(1)}_{agg} + (1-\lambda)R^{(2)}_{agg},\\
& \lambda \text{MSE}^{(1)}+(1-\lambda)\text{MSE}^{(2)}) \in \mathcal{C}.
\end{split}
\end{equation}
This result is shown using time-sharing arguments
(see Cover and Thomas~\cite{Cover06}).
Assume that $(T^{(1)},R^{(1)}_{agg},\text{MSE}^{(1)})$, $(T^{(2)},R^{(2)}_{agg},\text{MSE}^{(2)}) \in \mathcal{C}$ are achieved by probability distributions $\pi^{(1)}$ and $\pi^{(2)}$, respectively.
Let us select all parameters of the first tuple with probability
$\lambda$ and those of the second with probability $(1-\lambda)$.
Hence, we have
 $\pi = \lambda \pi^{(1)} + (1-\lambda) \pi^{(2)}$.
 Due to the linearity of expectation,
 $T = \lambda T^{(1)} + (1-\lambda) T^{(2)}$
 and $\text{MSE} = \lambda \text{MSE}^{(1)} + (1-\lambda) \text{MSE}^{(2)}$.
 Again, due to the linearity  of expectation,
 $R_{agg} = \lambda R^{(1)}_{agg} + (1-\lambda)R^{(2)}_{agg}$,
 implying that~\eqref{eq:suff1} is satisfied, and the proof is complete.
\end{IEEEproof}

\begin{myDef}\label{def:funcs}
{\em Let the function $R^*(T,\text{MSE}):\mathbb{R}_{\geq 0}^2\rightarrow \mathbb{R}_{\geq 0}$
be the Pareto optimal rate function, which is implicitly described as
$R^*(T,\text{MSE})=R_{agg}^* \Leftrightarrow (T,R_{agg}^*,\text{MSE}) \in \mathcal{P}$.
We further define implicit functions $T^*(R_{agg}, \text{MSE})$ and $\text{MSE}^*(T, R_{agg})$ in a similar way.}
\end{myDef}

\begin{myCoro}\label{coro:convexArguments}
The functions $R^*(T,\text{MSE})$, $T^*(R_{agg}, \text{MSE})$, and $\text{MSE}^*(T,R_{agg})$ are convex in their arguments.
\end{myCoro}

Note that our proof for the convexity of the set $\mathcal{C}$ might be extended to other iterative distributed learning algorithms
that transmit lossily compressed messages.

\subsection{Pareto optimal points via DP}

After proving that the achievable set $\mathcal{C}$ is convex, we apply  DP in Section~\ref{sec:DP}
to find the Pareto optimal points, and validate the convexity of the achievable set.

According to Definition~\ref{def:Pareto}, the resulting tuple $(T,$ $R_{agg},\text{MSE})$
computed using DP (Section~\ref{sec:DP}) is Pareto optimal on the discretized search spaces.
Hence, in this subsection, we run  DP to obtain the Pareto optimal points for a certain distributed linear system by sweeping the parameter $b$~\eqref{eq:cost}.

Consider the same setting as in Fig.~\ref{fig:RtAndEMSE}, except that we analyze MP platforms~\cite{pottie2000,estrin2002,EC2} for different $b$~\eqref{eq:cost}. Running the  DP scheme
of Section~\ref{sec:DP},
we obtain the optimal coding rate sequence $\mathbf{R}^*$ that yields the lowest combined cost while
providing a desired EMSE that is at most $\Delta\in \{1,2,...,5\}\times \text{MMSE}$ or equivalently
$\text{MSE}\in \{2,3,...,6\}\times \text{MMSE}$.
In Fig.~\ref{fig:rateVSiter}, we draw the Pareto optimal surface obtained by our DP scheme,
where the circles are Pareto optimal points.
Fig.~\ref{fig:Pareto2d_fixT} plots the aggregate coding rate $R_{agg}$
as a function of MSE for different optimal numbers of MP-AMP iterations $T$.
Finally, Fig.~\ref{fig:Pareto2d_fixMSE} plots the aggregate coding rate $R_{agg}$
as a function of $T$ for different optimal MSEs.
We can see that the surface comprised of the Pareto optimal points is indeed convex.
Note that when running DP to generate Fig.~\ref{fig:Pareto}, we used the  RD function~\cite{Cover06,Berger71,GershoGray1993,WeidmannVetterli2012} to model the
relation between the rate $R_t$ and distortion $D_t$ at each iteration,
which could be approached by VQ at sufficiently high
rates. We also ignored the constraint on the
quantization bin size (Section~\ref{sec:MP-CS_for_MP-AMP}).
Therefore, we only present Fig.~\ref{fig:Pareto} for illustration purposes.

When a smaller MSE  (or equivalently smaller EMSE)
is desired, more iterations $T$ and greater aggregate coding rates
$R_{agg}$~\eqref{eq:R_agg} are needed. Optimal coding rate sequences increase
$R_{agg}$ to reduce  $T$ when communication costs are
low (examples are commercial cloud computing systems~\cite{EC2}, multi-processor CPUs, and graphic processing units), whereas more iterations
allow to reduce the coding rate when communication is costly (for example, in sensor networks~\cite{pottie2000,estrin2002}).
These applications are discussed in Section~\ref{sec:realworld}.

{\bf Discussion of corner points:} We further discuss the
corners of the Pareto optimal surface (Fig.~\ref{fig:Pareto}) below.

\begin{enumerate}
\item First, consider the corner points along the MSE coordinate.
\begin{itemize}
\item If MSE$^* \rightarrow$ MMSE (or equivalently $\Delta \to 0$),
then MP-AMP needs to run infinite iterations with infinite coding rates. Hence, $R_{agg}^*\rightarrow \infty$ and $T^*\rightarrow \infty$. The rate of growth of $R_{agg}^*$ can be deduced from Theorem~\ref{th:optRateLinear}.
\item If MSE$^*$ $\rightarrow \rho$ (the variance of the signal~\eqref{eq:BG}), then MP-AMP does not need to run any iterations at all. Instead, MP-AMP outputs an all-zero estimate. Therefore, $\lim_{\text{MSE}^*\rightarrow\rho} R_{agg}^*= 0$ and $\lim_{\text{MSE}^*\rightarrow\rho} T^*=0$.
\end{itemize}
\item Next, we discuss the corner points along the $T$ coordinate.
\begin{itemize}
\item If $T^*\rightarrow 0$, then the best MP-AMP can do is to output an all-zero estimate. Hence, $\lim_{T^*\rightarrow 0} \text{MSE}^* = \rho$ and $\lim_{T^*\rightarrow 0} R_{agg}^*=0$.
\item The other extreme, $T^*\rightarrow \infty$, occurs only when
we want to achieve an MSE$^*\rightarrow$ MMSE.
Hence, $R_{agg}\rightarrow\infty$.
\end{itemize}
\item We conclude with corner points along the $R_{agg}$ coordinate.
\begin{itemize}
\item If $R_{agg}^*\rightarrow 0$, then the best MP-AMP can do is to output an all-zero estimate without running any iterations at all. Hence,
$\lim_{R_{agg}^*\rightarrow 0}\text{MSE}^*=\rho$ and $\lim_{R_{agg}^*\rightarrow 0} T^* = 0$.
\item If $R_{agg}^*\rightarrow\infty$, then the optimal scheme will use high rates in all iterations,
and  MP-AMP resembles centralized AMP. Therefore, the MSE$^*$ as a function of $T^*$ converges to
that of centralized AMP SE~\eqref{eq:ori_SE}.
\end{itemize}
\end{enumerate}

\section{Real-world Case Study}\label{sec:realworld}

To showcase the difference between optimal coding rate sequences in different platforms,
this section discusses several MP platforms including
sensor networks~\cite{pottie2000,estrin2002}
and large-scale cloud servers~\cite{EC2}. The costs in these platforms are
quite different due to the different constraints in these platforms, and we will see how they affect the optimal coding rate sequence $\mathbf{R}^*$.
The changes in the optimal $\mathbf{R}^*$ highlight the importance of optimizing for the correct costs.

\subsection{Sensor networks}\label{sec:sn}
In sensor networks~\cite{pottie2000,estrin2002}, distributed sensors are typically dispatched to remote locations where they collect data and communicate with the fusion center. However, distributed sensors may have severe power consumption constraints. Therefore, low power chips such as the CC253X from Texas Instruments~\cite{CC2530} are commonly used in distributed sensors. Some typical parameters for such low power chips are: central processing unit (CPU) clock frequency 32MHz, data transmission rate 250Kbps, voltage between 2V-3.6V, and transceiver current 25mA~\cite{CC2530}, where the CPU current resembles the transceiver current. Because these chips are generally designed to be low power, when transmitting and receiving data, the CPU helps the transceiver and cannot carry out computing tasks. Therefore, the power consumption can be viewed as constant. Hence, in order to minimize the power consumption, we minimize the total runtime when estimating a signal from MP-LS measurements~\eqref{eq:one-node-meas} collected by the distributed sensors.

The runtime in each MP-AMP iteration~\eqref{eq:slave1}-\eqref{eq:master} consists of ({\em i}) time for computing~\eqref{eq:slave1} and~\eqref{eq:slave2}, ({\em ii}) time for encoding $\f_t^p$~\eqref{eq:slave2}, and ({\em iii}) data transmission time for $Q(\f_t^p)$~\eqref{eq:master0}.
As discussed in Section~\ref{sec:MP-CS_for_MP-AMP}, the fusion center may broadcast $\x_t$~\eqref{eq:master}, and simple compression schemes can reduce the coding rate. Therefore, we consider the data reception time in the $P$ processor nodes to be constant.
The overall computational complexity for~\eqref{eq:slave1} and~\eqref{eq:slave2} is $O(\frac{MN}{P})$.
Suppose further that ({\em i}) each processor node needs to carry out two matrix-vector
products in each iteration, ({\em ii}) the overhead of moving data in memory is assumed to
be 10 times greater than the actual computation, and ({\em iii}) the clock frequency is 32MHz.
Hence, we assume that the actual time needed for computing~\eqref{eq:slave1} and~\eqref{eq:slave2}
is $C_4=\frac{20MN}{32\times 10^6 P}$ sec. Transmitting $Q(\f_t^p)$ of length $N$ at coding rate
$R$ requires $\frac{RN}{250\times 10^3}$ sec, where the denominator is the data transmission rate
of the transceiver. Assuming that the overhead in communication is approximately the same as the
communication load caused by  the actual messages, we obtain that the time requested for
transmitting $Q(\f_t^p)$ at coding rate $R$ is $C_5 R$ sec, where $C_5=\frac{2N}{250\times 10^3}$.
Therefore, the total cost can be calculated from~\eqref{eq:cost} with $b=\frac{C_4}{C_5}$~\eqref{eq:cost}.

Because low power chips equipped in distributed sensors have limited memory (around 10KB, although sometimes external flash is allowed)~\cite{CC2530}, the signal length $N$ and number of measurements $M$ cannot be too large. We consider $N=1,\!000$ and $M=400$ spread over $P=100$ sensors, sparsity rate $\rho=0.1$, and $\sigma_Z^2=\frac{1}{400}$.
We set the desired MSE to be $0.5$ dB above the MMSE, i.e., $10\log_{10}\l(1+ \frac{\Delta}{\text{MMSE}}\r)=0.5$,
and run DP as in
Section~\ref{sec:DP}.\footnote{Throughout Section~\ref{sec:realworld}, we use the RD function~\cite{Cover06,Berger71,GershoGray1993,WeidmannVetterli2012} to model the relation between
rate $R_t$ and distortion $D_t$ at each iteration. We also ignore the constraint on the quantizer
(Section~\ref{sec:MP-CS_for_MP-AMP}). Therefore, the optimal coding rate sequences in
Section~\ref{sec:realworld} are only for illustration purposes.
}
The coding rate sequence provided by DP is $\mathbf{R}^*=(0.1$, $0.1$, $0.6$, $0.8$, $1.0$, $1.0$, $1.1$, $1.1$, $1.2$, $1.4$, $1.6$, $1.9$, $2.3$, $2.7$, $3.1)$. In total we have $T=15$ MP-AMP iterations with $R_{agg}=20.0$ bits aggregate coding rate~\eqref{eq:R_agg}.
The final MSE ($\text{MMSE} + \Delta$) is $7.047\times 10^{-4}$, which is 0.5 dB from the MMSE ($6.281\times 10^{-4}$)~\cite{ZhuBaronCISS2013,Krzakala2012probabilistic,GuoBaronShamai2009,RFG2012}.

\iftoggle{TSP}{}{
\begin{figure*}[t]
\centering
\subfigure[]{\label{fig:PearsonWN}
\includegraphics[width=0.48\textwidth]{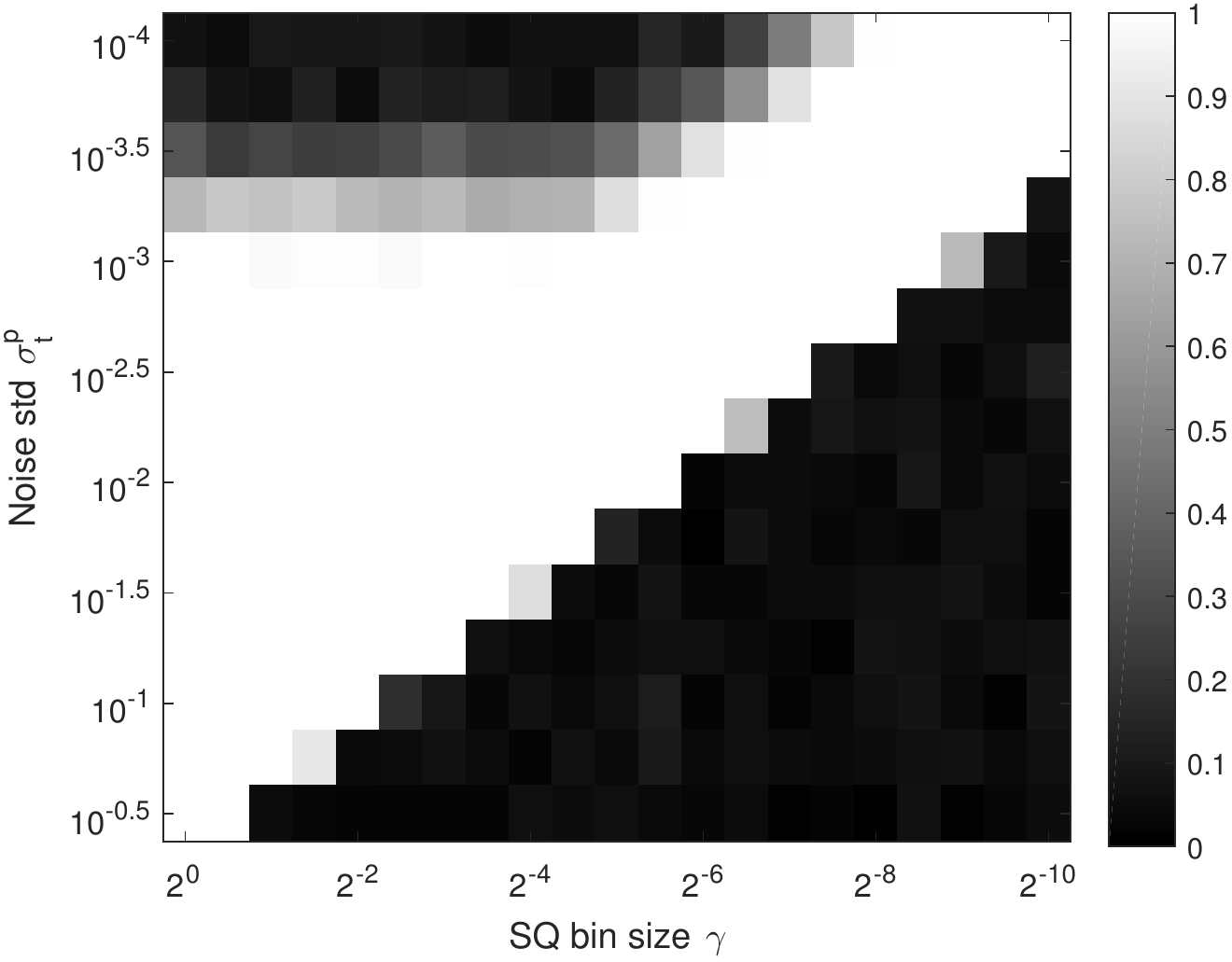}}
\subfigure[]{\label{fig:PearsonWpNX}
\includegraphics[width=0.48\textwidth]{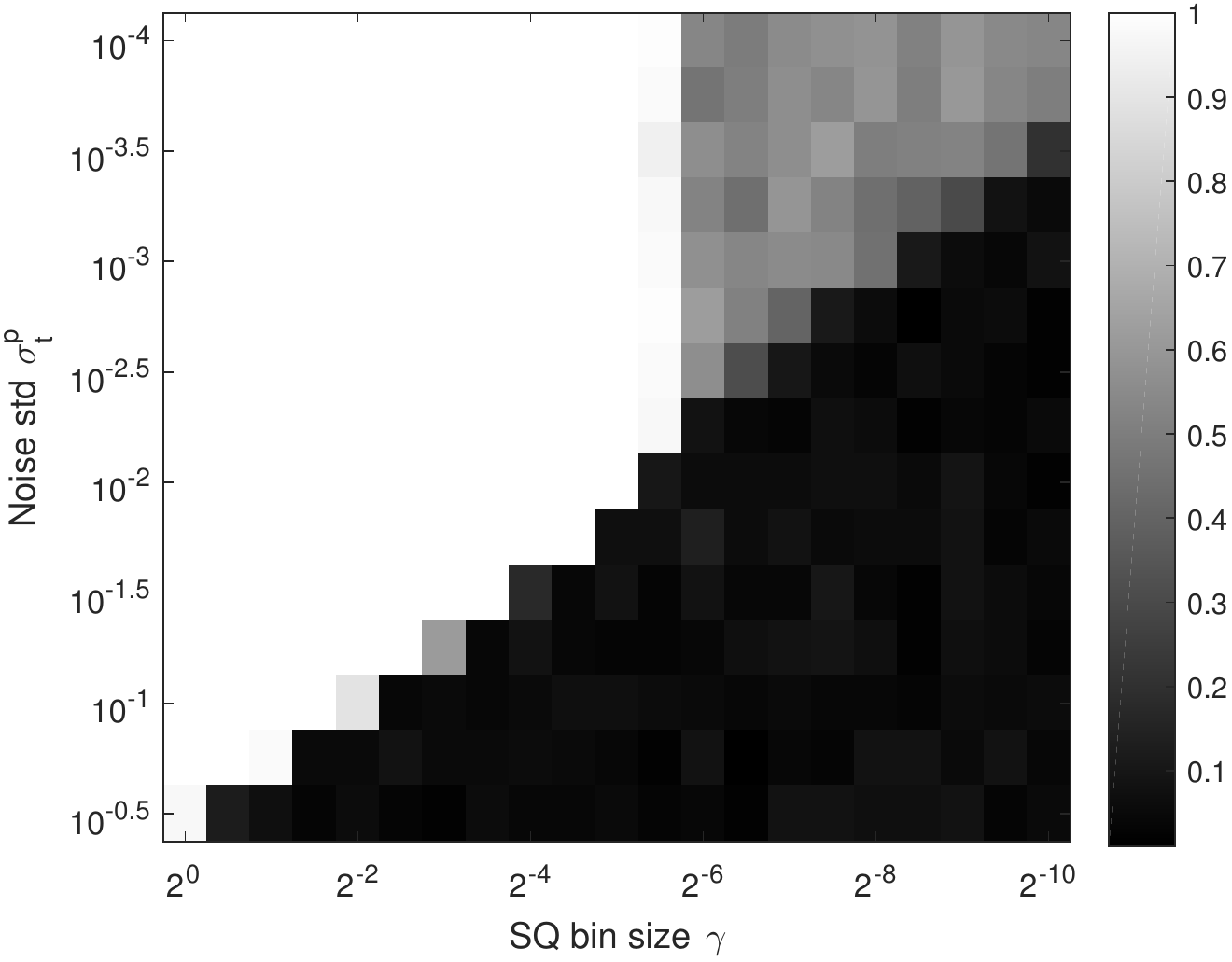}}
\caption{PCC test results. The shades of gray show the fraction of
100 tests where we reject the null hypothesis (random variables being tested are uncorrelated)
with 5\% confidence. The horizontal and vertical axes represent the quantization bin size $\gamma$
of the SQ and the scalar channel noise standard deviation (std) $\sigma_t^p$ in each processor node,
respectively. Panel (a): Test the correlation between $\w_t$ and $\n_t$.
Panel (b): Test the correlation between $\w_t+\n_t$ and $\x$.}
\end{figure*}
}

\subsection{Large-scale cloud server}\label{sec:largeScaleCase}
Having discussed sensor networks~\cite{pottie2000,estrin2002}, we now discuss an application of  DP (cf. Section~\ref{sec:DP}) to large-scale cloud servers. Consider the dollar cost for users of Amazon EC2~\cite{EC2}, a commercial cloud computing service. A typical cost for CPU time is $\$ 0.03$/hour, and the data transmission cost is $\$ 0.03$/GB. Assuming that the CPU clock frequency is 2.0GHz and considering various overheads, we need a runtime of $\frac{20MN}{2\times 10^9 P}$ sec and the computation cost is $C_4=\$ \frac{20MN}{2\times 10^9 P}\times \frac{0.03}{3600}$ per MP-AMP iteration.
Similar to Section~\ref{sec:sn}, the communication cost for coding rate $R$ is $C_5 R=\$ 2R N\frac{0.03}{8\times 10^9}$.
Note that the multiplicative factors of 20 in $C_4$ and 2 in $C_5$ are due
to the same considerations as in Section~\ref{sec:sn}, and the $8\times 10^9$ in $C_5$ is the number of bits per GB.
Therefore, the total cost with $T$ MP-AMP iterations can still be modeled as in~\eqref{eq:cost}, where $b=\frac{C_4}{C_5}$.

We consider a problem with the same signal and channel model as the setting of Section~\ref{sec:sn}, while the size of the problem grows to $N=50,\!000$ and $M=20,\!000$ spread over $P=100$ computing nodes. Running DP, we obtain the coding rate sequence $\mathbf{R}^*=(1.3$, $1.6$, $1.8$, $1.8$, $1.8$, $1.9$, $2.1$, $2.3$, $2.6$, $3.1$, $3.7)$ for a total of $T=11$ MP-AMP iterations with $R_{agg}=24.0$ bits aggregate coding rate. The final MSE is $7.031\times 10^{-4}$, which is 0.49 dB
above the MMSE. Note that this final MSE is 0.01 dB better than our goal of
$0.5$ dB above the MMSE due to the discretized search spaces used in DP.

{\bf Settings with even cheaper communication costs:} Compared to large-scale cloud servers, the relative
cost of communication is even cheaper in multi-processor CPU and graphics processing unit (GPU)
systems. We reduce $b$ by a factor of 100 compared to the large-scale cloud server case above. We rerun DP, and obtain the
coding rate sequence $\mathbf{R}^*=(2.3$, $2.5$, $2.6$, $2.7$, $2.7$, $2.8$, $3.0$, $3.4$, $3.7$, $4.5)$ for $T=10$ and $R_{agg}=30.2$ bits. Note that 10 iterations are needed for centralized AMP to converge in this setting. With the low-cost communication of this setting, DP yields a coding rate sequence $\mathbf{R}^*$ within 0.5 dB of the MMSE with the same number of iterations as centralized AMP, while using an average coding rate of only 3.02 bits  per iteration.

\begin{myRemark}
Let us review the cost tuples $(T,R_{agg},\text{MSE})$ for our three cases. For sensor networks, $(T,R_{agg},\text{MSE})_{\text{sensornet}}=(15,20,7.047\times 10^{-4})$; for cloud servers,
$(T,R_{agg},\text{MSE})_{\text{cloud}}$ $=$ $(11,24,7.031\times 10^{-4})$; and for GPUs, $(T,R_{agg},\text{MSE})_{\text{GPU}}=(10,30.2,7.047\times 10^{-4})$. These cost tuples are different points in the Pareto optimal set $\mathcal{P}$~\eqref{eq:setP}.
We can see for sensor networks that the optimal coding rate sequence
reduces $R_{agg}$ while adding
iterations, because sensor networks have relatively expensive communications.
The optimal coding rate sequences use higher rates in cloud servers and GPUs,
because their communication costs are relatively lower.
Indeed, different trade-offs between computation and communication
lead to different aggregate coding rates $R_{agg}$ and numbers of MP-AMP iterations $T$.
Moreover, the optimal coding rate sequences for sensor networks, cloud servers, and GPUs
use average coding rates of $1.33$, $2.18$, and $3.02$ bits/entry/iteration, respectively.
Compared to $32$ bits/entry/iteration single-precision floating point communication
schemes, optimal coding rate sequences reduce the communication costs significantly.
\end{myRemark}

\section{Conclusion}\label{sec:conclude}
This paper used lossy compression in multi-processor (MP) approximate message passing (AMP) for solving MP linear inverse problems. Dynamic programming (DP) was used to obtain the optimal coding rate sequence for MP-AMP that incurs the lowest combined cost of communication and computation while achieving a desired
mean squared error (MSE). We posed the problem of finding the optimal coding rate sequence in the low excess MSE (EMSE=MSE-MMSE, where MMSE refers to the minimum MSE) limit as a convex optimization problem and proved that optimal coding rate sequences are approximately linear when the EMSE is small.
Additionally, we obtained that the combined cost of computation
and communication scales with $O(\log^2(1/\text{EMSE}))$.
Furthermore, realizing that there is a trade-off among the communication cost, computation cost, and MSE, we formulated a multi-objective optimization problem (MOP) for these costs and studied the Pareto optimal points that exploit this trade-off. We proved that the achievable region of the MOP is convex.

We further emphasize that there is little work in the prior art discussing the optimization of communication schemes in iterative distributed algorithms. Although we focused on the MP-AMP algorithm, our conclusions such as the linearity of the optimal coding rate sequence and the convexity of the
achievable set of communication/computation trade-offs could be extended to
other iterative distributed algorithms including consensus averaging~\cite{Frasca2008,Thanou2013}.

\section*{Acknowledgments}
The authors thank Puxiao Han and Ruixin Niu for numerous discussions about MP settings of linear systems and AMP.  Mohsen Sardari and Mihail Sichitiu offered useful information about the practical distributed computing platforms of Section~\ref{sec:realworld}, and Cynthia Rush provided useful insights about the proof of AMP SE. The authors also thank Yanting Ma and Ryan Pilgrim for helpful comments about the manuscript.

\appendix

\iftoggle{TSP}{}{
\subsection{Impact of the quantization error}\label{app:verifyIndpt}

This appendix provides numerical evidence that ({\em i}) the quantization error $\n_t$
is independent of the scalar channel noise $\w_t$~\eqref{eq:indpt_noises} in the fusion center and
({\em ii}) $\w_t+\n_t$ is independent of the signal $\x$. In the following, we simulate the AMP equivalent scalar channel in each processor node and in the fusion center.
In the interest of simple implementation, we use scalar quantization (SQ)
to quantize $\f_t^p$~\eqref{eq:slave2} (in each processor node) and
hypothesis testing to evaluate ({\em i}) whether $\w_t$ and $\n_t$ (in the fusion center) are independent and ({\em ii}) whether $\w_t+\n_t$ and $\x$ are independent. Both parts are necessary for lossy SE~\eqref{eq:SE_Q} to hold: part ({\em i}) ensures that we can predict the variance of $\w_t+\n_t$ by $\sigma_t^2+PD_t$ and part ({\em ii}) ensures that lossy MP-AMP falls within the general framework of Bayati and Montanari~\cite{Bayati2011} and Rush and Venkataramanan~\cite{Rush_ISIT2016_arxiv}, so that lossy SE~\eqref{eq:SE_Q} holds. Details about our simulation appear below.

\begin{figure*}[t]
\centering
\subfigure[]{\label{fig:lossySEvsSim_BG}
\includegraphics[width=0.48\textwidth]{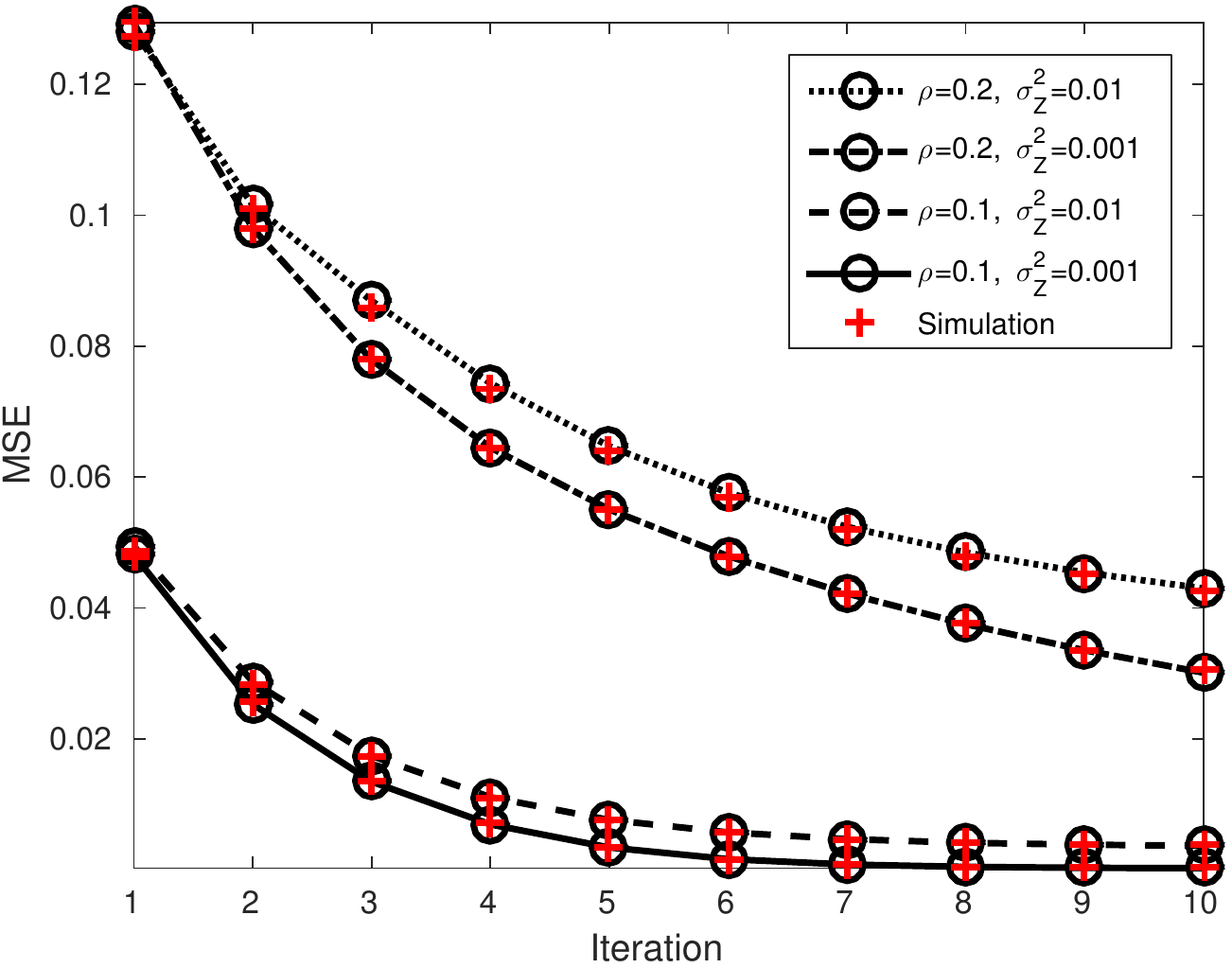}}
\subfigure[]{\label{fig:lossySEvsSim_mixG}
\includegraphics[width=0.48\textwidth]{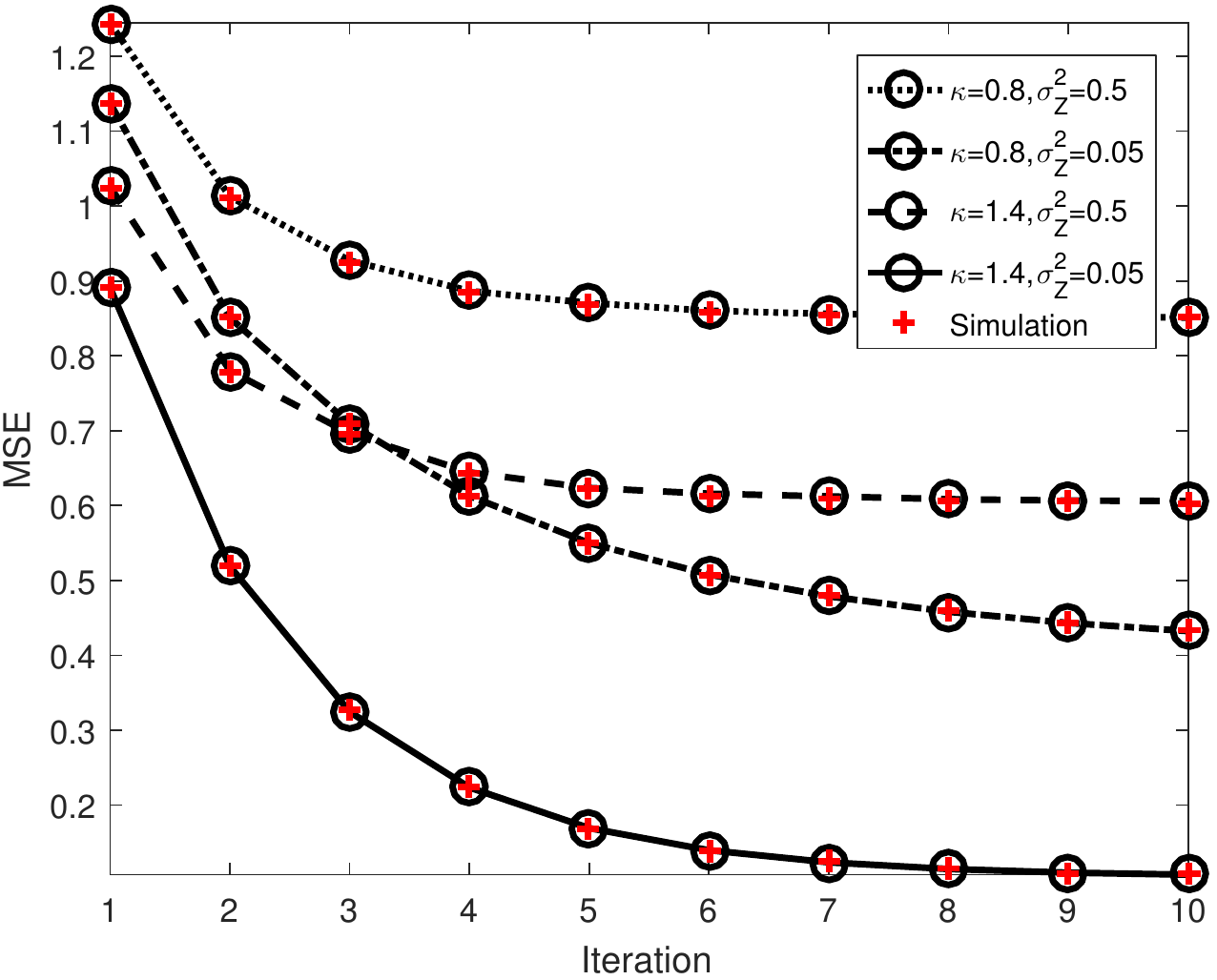}}
\caption{Comparison of the MSE predicted by lossy SE~\eqref{eq:SE_Q}
and the MSE of MP-AMP simulations for various settings.
The round markers represent MSEs predicted by lossy SE, and the (red) crosses represent
simulated MSEs. Panel (a): Bernoulli Gaussian signal. Panel (b): Mixture Gaussian signal.}\label{fig:lossySEvsSim}
\end{figure*}

Considering~\eqref{eq:equivalent_scalar_channel} and~\eqref{eq:slave2}, we obtain that the AMP equivalent scalar channel in each processor node can be expressed as
\begin{equation}\label{eq:noisePseudoNode}
\f_t^p=\frac{1}{P}\x+\w_t^p,
\end{equation}
where $\sum_{p=1}^P \w_t^p=\w_t$~\eqref{eq:equivalent_scalar_channel}, 
and the variances of $\w_t^p$ and $\w_t$ can be expressed as $(\sigma_t^p)^2$ and $\sigma_t^2$, respectively~\eqref{eq:SE_Q}. Hence, we obtain $\sigma_t^2=\sum_{p=1}^P (\sigma_t^p)^2$. The signal $\x$ follows~\eqref{eq:BG} with $\rho=0.1$. The entries of $\w_t^p$ are i.i.d. and follow $\mathcal{N}(0,(\sigma_t^p)^2)$. Next, we apply an SQ to $\f_t^p$~\eqref{eq:noisePseudoNode},
\begin{equation}\label{eq:quantPseudoNode}
Q(\f_t^p)=\frac{1}{P}\x+\w_t^p+\n_t^p,
\end{equation}
where $Q(\cdot)$ denotes the quantization process, $\n_t^p$ is the quantization error
in processor node $p$, and recall that the variance of $\n_t^p$ is $D_t$.
We simulate the fusion center by calculating
\begin{equation}\label{eq:processInfusionCenter}
\f_t=\sum_{p=1}^P Q(\f_t^p)=\x+\w_t+\n_t,
\end{equation}
where $\n_t=\sum_{p=1}^P \n_t^p$.
Note that $\w_t$ is Gaussian due to properties of AMP~\cite{DMM2009,Montanari2012,Bayati2011}.
The total quantization error at the fusion center, $\n_t$, is also Gaussian,
due to the central limit theorem. Hence, in order to test the independence
of $\w_t$ and $\n_t$~\eqref{eq:processInfusionCenter}, we need only test whether
$\w_t$ and $\n_t$ are uncorrelated. We also test whether $\w_t+ \n_t$ and $\x$ are uncorrelated.

We study the settings $\sigma_t^p\in\{10^{-0.5},\cdots,10^{-4}\}$ and $\gamma\in\{2^{0},\cdots,2^{-10}\}$,
where $\gamma$ denotes the SQ bin size. In each setting,
we simulate~\eqref{eq:noisePseudoNode}--\eqref{eq:processInfusionCenter} 100 times and
perform 100 Pearson correlation coefficient (PCC) tests~\cite{PCC} for $\w_t$ and $\n_t$,
respectively. The null hypothesis of the PCC tests~\cite{PCC}
is that $\w_t$ and $\n_t$ are uncorrelated. The null hypothesis is rejected
if the resulting $p$-value is smaller than 0.05.

For each setting, we record the fraction of 100 tests where
the null hypothesis is rejected, which is shown by the shades of gray
in Fig.~\ref{fig:PearsonWN}. The horizontal and vertical axes represent the
quantization bin size $\gamma$ and the standard deviation (std) $\sigma_t^p$,
respectively. Similarly, we test $\w_t+\n_t$ and $\x$; results appear in
Fig.~\ref{fig:PearsonWpNX}. We can see that when $\gamma \ll \sigma_t^p$
(bottom right corner),
({\em i}) $\w_t$ and $\n_t$ tend to be independent and
({\em ii}) $\w_t+\n_t$ and $\x$ tend to be independent.

Now consider Fig.~\ref{fig:PearsonWpNX}, which provides PCC test results
evaluating possible correlations between $\w_t+\n_t$ and $\x$.
There appears to be a phase transition that separates regions where $\w_t+\n_t$
and $\x$ seem independent or dependent. We speculate that this phase transition
is related to the pdf of $\frac{1}{P}\x+\w_t^p$. To explain our hypothesis,
note that when the noise $\w_t^p$ is low (top part of~Fig.~\ref{fig:PearsonWpNX}),
the phase transition is less affected by noise, and the role of $\gamma$
is smaller. By contrast, large noise (bottom) sharpens the phase transition.

In summary, it appears that when $\gamma<2\sigma_t^p=\frac{2\sigma_t}{\sqrt{P}}$,
we can regard ({\em i}) $\w_t$ and $\n_t$ to be independent and
({\em ii}) $\w_t+\n_t$ and $\x$ to be independent. The requirement
$\gamma<2\sigma_t^p=\frac{2\sigma_t}{\sqrt{P}}$ is motivated by Widrow and
Koll{\'a}r~\cite{widrow2008quantization};
we leave the study of this phase transition for future work.

\subsection{Numerical evidence for lossy SE}\label{app:verifyLossySE}
This appendix provides numerical evidence for lossy SE~\eqref{eq:lossySE}.
We simulate two signal types, one is the Bernoulli Gaussian signal~\eqref{eq:BG}
and the other is a mixture Gaussian.

{\bf Bernoulli Gaussian signals:} We generate 50 signals of length $10,\!000$
according to~\eqref{eq:BG}. These signals are measured by $M=5,\!000$ measurements
spread over $P=100$ distributed nodes. We estimate each of these signals by running
$T=10$ MP-AMP iterations. ECSQ is used to quantize $\f_t^p$~\eqref{eq:noisePseudoNode},
and $Q(\f_t^p)$~\eqref{eq:quantPseudoNode} is encoded at coding rate $R_t$.
We simulate settings with sparsity rate $\rho\in\{0.1,0.2\}$
and noise variance $\sigma_Z^2\in\{0.01,0.001\}$.
In each setting, we randomly generate the coding rate sequence $\mathbf{R}$,
s.t. the quantization bin size at each iteration satisfies $\gamma<\frac{2\sigma_t}{\sqrt{P}}$
(details in Appendix~\ref{app:verifyIndpt}).\footnote{Note that the constraint on $\gamma$
implies that $\mathbf{R}$ is likely monotone non-decreasing.}
A Bayesian denoiser,
$\eta_t(\cdot)=\mathbb{E}[\x|\f_t]$, is used in~\eqref{eq:master}.
The resulting MSEs from the MP-AMP simulation averaged over the 50 signals,
along with MSEs predicted by lossy SE~\eqref{eq:lossySE},
are plotted in Fig.~\ref{fig:lossySEvsSim_BG}.
We can see that the simulated MSEs are close to the MSEs predicted by lossy SE.

{\bf Mixture Gaussian signals:} We independently generate 50 signals of length $10,\!000$ according to
$X = \sum_{i\in \{0,1,2\}} \mathbbm{1}_{X_B = i} X_{G,i}$
where $X_B \sim \text{cat}(0.5, 0.3, 0.2)$ follows a categorical distribution on alphabet $\{0,1,2\}$, $X_{G,0} \sim \mathcal{N}(0,0.1)$, $X_{G,1} \sim \mathcal{N}(-1.5,0.8)$, and $X_{G,2} \sim \mathcal{N}(2,1)$.
We simulate settings with $T=10$, $P=100$, $\kappa=\frac{M}{N}\in\{0.8,1.6\}$,
and $\sigma_Z^2\in\{0.5,0.05\}$. In each setting, we randomly generate the coding rate  sequence $\mathbf{R}$, s.t. the quantization bin size at each iteration satisfies $\gamma<\frac{2\sigma_t}{\sqrt{P}}$. The results are plotted in Fig.~\ref{fig:lossySEvsSim_mixG}. The simulation results match well with the lossy SE predictions.

\subsection{Integrity of discretized search space}\label{app:Integrity}

\begin{figure}
\centering
 \includegraphics[width=0.48\textwidth]{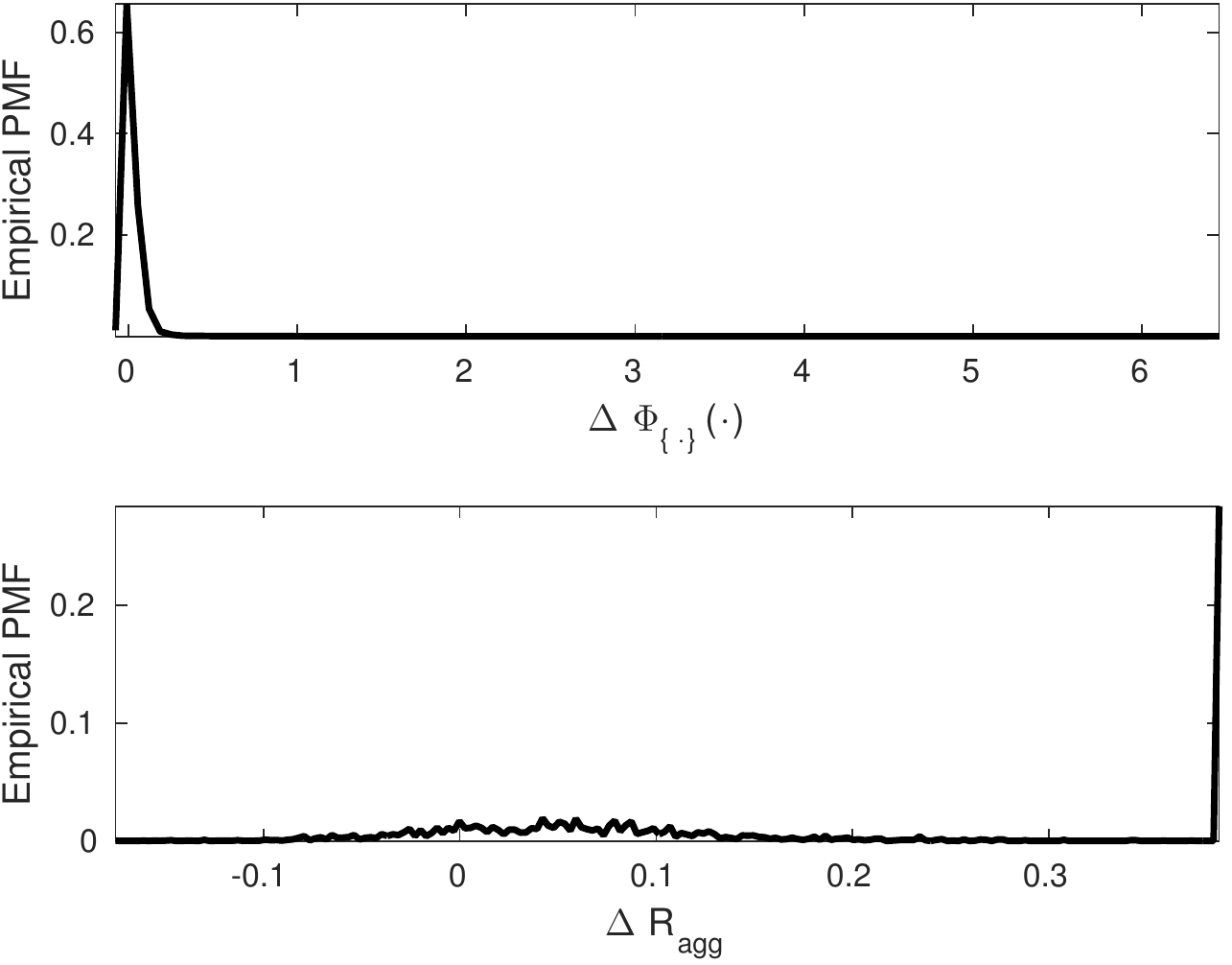}
  \caption{Justification of the discretized search space used in DP. Top panel:
Empirical PMF of the error in the cost function $\Delta \Phi_{\{\cdot\}}(\cdot)$ used to verify the integrity of the linear interpolation in the discretized search space of $\sigma^2$. Bottom panel:
Empirical PMF of $\Delta R_{agg}$; used to verify the integrity of the choice of $\Delta R=0.1$.}\label{fig:DPintegrity}
\end{figure}

When a coding rate $\widehat{R}$ is selected in MP-AMP iteration $t$, DP calculates the equivalent scalar channel noise variance $\sigma^2_{t+1}$~\eqref{eq:indpt_noises} for the next MP-AMP iteration according to~\eqref{eq:SE_Q}. The variance $\sigma^2_{t+1}$ is unlikely to lie on the discretized search space for $\sigma_t^2$, denoted by the {\em grid} $\mathcal{G}(\sigma^2)$. Therefore, $\Phi_{T-(t+1)}(\sigma^2_{t+1}(\widehat{R}))$ in~\eqref{eq:DPrecursion} does not reside in memory.
Instead of brute-force calculation of
$\Phi_{\{\cdot\}}(\cdot)$, we estimate it by fitting a function to the closest neighbors of $\sigma^2_{t+1}$ that lie on the grid $\mathcal{G}(\sigma^2)$  and finding $\Phi_{\{\cdot\}}(\cdot)$ according to the fit function. We evaluate a linear interpolation scheme.

{\bf Interpolation in $\mathcal{G}(\sigma^2)$:}
We run DP over the original coarse grid $\mathcal{G}^c(\sigma^2)$ with resolution $\Delta \sigma^2=0.01$ dB, and a 4$\times$ finer grid $\mathcal{G}^f(\sigma^2)$ with  $\Delta \sigma^2=0.0025$ dB. We obtain the cost function with the coarse grid $\Phi^c_{T-t} (( \sigma^2_t)_c)$ and the cost function with the fine grid $\Phi^f_{T-t} ((\sigma^2_t)_f),\ \forall t\in\{1,...,T\}, (\sigma^2_t)_c\in \mathcal{G}^c(\sigma^2), (\sigma^2_t)_f\in \mathcal{G}^f(\sigma^2)$. Next, we interpolate $\Phi^c_{T-t} (( \sigma^2_t)_c)$ over the fine grid $\mathcal{G}^f(\sigma^2)$ and obtain the interpolated $\Phi^i_{T-t} (( \sigma^2_t)_c)$.
In order to compare $\Phi^i_{T-t} (( \sigma^2_t)_c)$ with $\Phi^f_{T-t} (( \sigma^2_t)_c)$ in a comprehensive way, we consider the settings given by the Cartesian product of the following variables: ({\em i}) the number of distributed nodes $P\in\{50,100\}$, ({\em ii}) sparsity rate $\rho\in\{0.1,0.2\}$, ({\em iii}) measurement rate $\kappa=\frac{M}{N}\in\{3\rho,5\rho\}$, ({\em iv}) EMSE $\epsilon_T \in \{1,0.5\}$dB, ({\em v}) parameter $b\in\{0.5,2\}$, and ({\em vi}) noise variance $\sigma_Z^2\in \{0.01,0.001\}$.
In total, there are 64 different settings.
We calculate the error $\Delta\Phi_{T-t}\l(( \sigma^2_t)_c\r)=\Phi^i_{T-t} \l(( \sigma^2_t)_c\r)-\Phi^f_{T-t} \l(( \sigma^2_t)_c\r)$ and plot the empirical probability mass function (PMF) of $\Delta\Phi_{T-t}\l(( \sigma^2_t)_c\r)$
over all $t$, $(\sigma^2_t)_c$, and all 64 settings. The resulting empirical PMF
of $\Delta\Phi_{\{\cdot\}}(\cdot)$ is plotted in the top panel of Fig.~\ref{fig:DPintegrity}.
We see that with 99\% probability, the error satisfies $\Delta\Phi_{\{\cdot\}}\l(\cdot\r)\leq 0.2$,
which corresponds to an inaccuracy of approximately 0.2 in the aggregate
coding rate $R_{agg}$.\footnote{Note that when calculating $\Phi^f$, we are still
using the corresponding interpolation scheme. Although this comparison is not ideal,
we believe it still provides the reader with enough insight.} In the simulation,
we used a resolution of $\Delta R=0.1$. Hence, the inaccuracy of 0.2 in $R_{agg}$
(over roughly 10 iterations) is negligible. Therefore, we use linear interpolation
with a coarse grid $\mathcal{G}^c(\sigma^2)$ with  $\Delta \sigma^2=0.01$ dB.

{\bf Integrity of choice of $\Delta R$:}
We tentatively select resolution $\Delta R=0.1$, and investigate the integrity of this $\Delta R$  over the 64 different settings above. After the coding rate sequence $\mathbf{R}^*=(R_1^*,\cdots,R_T^*)$ is obtained by DP for each setting, we randomly perturb $R_t^*$ by $R_p(t)=R_t^*+\beta_t,\ t=1,...,T$,
where $R_p(t)$ is the {\em perturbed coding rate},
the bias is $\beta_t\in \left[-\frac{\Delta R}{2},+\frac{\Delta R}{2}\right]$,
and $\mathbf{R}_p=(R_p(1),\cdots,R_p(T))$ is called the {\em perturbed coding rate sequence}.
After randomly generating 100 different perturbed coding rate sequences $\mathbf{R}_p$,
we calculate the aggregate coding rate~\eqref{eq:R_agg}, $R_{agg}^p$, of each $\mathbf{R}_p$;
we only consider the perturbed coding rate sequences that achieve EMSE no greater
than the optimal coding rate sequence $\mathbf{R}^*$ given by DP. The bottom panel
of Fig.~\ref{fig:DPintegrity} plots the empirical PMF of $\Delta R_{agg}$, where
$\Delta R_{agg}=R_{agg}^p-R_{agg}^*$ and $R_{agg}^*=||\mathbf{R}^*||_1$.
Roughly 15\% of cases in our simulation yield $\Delta R_{agg}<0$
(meaning that the perturbed coding rate sequence has lower $R_{agg}$),
while for the other 85\% cases, $\mathbf{R}^*$ has lower $R_{agg}$.
Considering the resolution $\Delta R=0.1$, we can see that the perturbed
sequences are only marginally better than $\mathbf{R}^*$.
Hence, we verified the integrity of $\Delta R=0.1$.
}

\subsection{Proof of Lemma~\ref{lemma:DPoptimal}}\label{app:proofDPoptimal}
\begin{proof}
We show that our DP scheme~\eqref{eq:DPrecursion} fits into Bertsekas' formulation ~\cite{bertsekas1995},
which has been proved to be optimal.
Under Bertsekas' formulation, our decision variable is the coding rate $R_t$ and our state
is the scalar channel noise variance $\sigma_t^2$. Our next-state function is the lossy
SE~\eqref{eq:SE_Q} with the distortion $D_t$ being calculated from the RD function given the decision variable $R_t$.
Our additive cost associated with the dynamic system is $b\times \mathbbm{1}_{R_t\neq 0}+ R_t$.
Our control law maps the state $\sigma_t^2$ to a decision (the coding rate $R_t$).
Therefore, our DP formulation~\eqref{eq:DPrecursion} fits into the optimal DP formulation of
Bertsekas~\cite{bertsekas1995}. Hence, our DP formulation~\eqref{eq:DPrecursion}
is also optimal {\em for the discretized search spaces of $R_t$ and $\sigma_t^2$}.
\end{proof}

\subsection{Proof of Theorem~\ref{th:optRateLinear}}\label{app:optRateLinear}
\begin{IEEEproof}
Our proof is based on the assumption that lossy SE~\eqref{eq:SE_Q} holds.
Consider the geometry of the SE incurred by $\mathbf{R}^*$ for arbitrary iterations $t$ and $t+1$, as shown in Fig.~\ref{fig:evolveEMSE}.
Let $\widetilde{S}_t = (\widetilde{\sigma}_t^2,\widetilde{u}_t)$ and $R_t^*$ be the state and the optimal coding rate at iteration $t$, respectively.
We know that the slope of $\widetilde{g}_I(\cdot)$ is $\widetilde{g}_I'(\cdot)=1$. Hence,
the length of line segment $\widetilde{M}_t\widetilde{I}_t$  is $\widetilde{\sigma}_{t+1}^2=\widetilde{u}_t+PD_t$. That is
\begin{equation}\label{eq:iter_t}
PD_t=\widetilde{\sigma}_{t+1}^2-\widetilde{u}_t.
\end{equation}
Similarly, we obtain
\begin{equation}\label{eq:iter_tp1}
PD_{t+1}=\widetilde{\sigma}_{t+2}^2-\widetilde{u}_{t+1},
\end{equation}
where $\widetilde{u}_{t+1}$ and $\widetilde{\sigma}_{t+1}^2$ obey
\begin{equation}\label{eq:relate_2iters}
\widetilde{\sigma}_{t+1}^2=\widetilde{g}_S^{-1}(\widetilde{u}_{t+1}).
\end{equation}
Recall that, according to Taylor's theorem~\eqref{eq:Taylor_inverse}, we obtain that
\begin{equation}\label{eq:TaylorInverse}
\widetilde{g}_S^{-1}(\widetilde{u}_{t+1})=\frac{1}{\theta}\widetilde{u}_{t+1}+C \widetilde{u}_{t+1}^2,
\end{equation}
with $\theta$ defined in~\eqref{eq:theta}.
Although $C$ depends on $\widetilde{u}_{t+1}$, it is uniformly bounded, i.e., $C \in [-B,B]$ for some $0\leq B< \infty$.

\begin{figure}[t]
\centering
\includegraphics[width=0.48\textwidth]{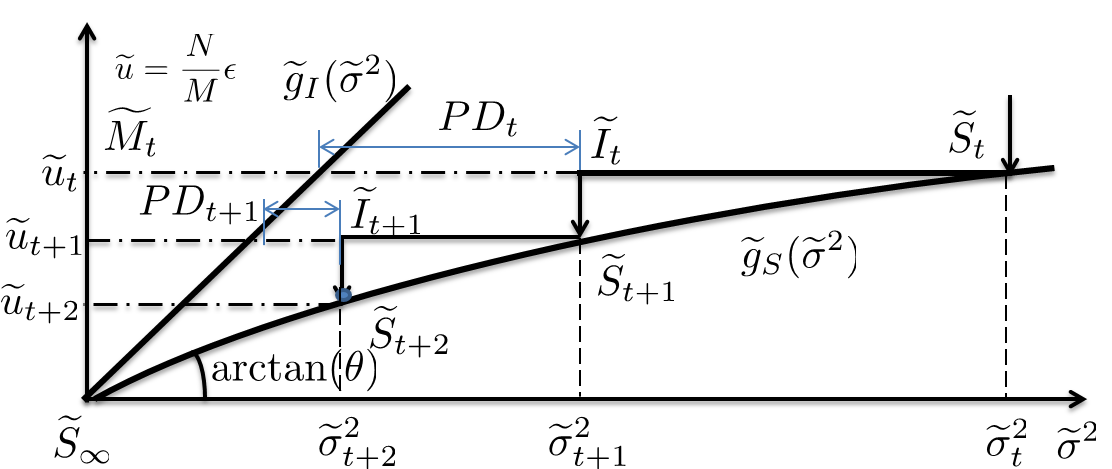}
\caption{Illustration of the evolution of $u_t$. The vertical axis shows $u_t=\frac{N}{M}\text{EMSE}=\frac{N}{M}\epsilon_t$.  The solid lines with arrows denote the lossy SE associated with a coding rate sequence and dashed-dotted lines are auxiliary lines.}\label{fig:evolveEMSE}
\end{figure}

Fixing $\widetilde{u}_{t}=\frac{N}{M}\epsilon_{t}^*$ and $\widetilde{u}_{t+2}=\frac{N}{M}\epsilon_{t+2}^*$, we  explore different distortions $D_{t}$ and $D_{t+1}$ that obey~\eqref{eq:iter_t}--\eqref{eq:relate_2iters}. According to Definition~\ref{def:optRate}, among distortions that obey~\eqref{eq:iter_t}--\eqref{eq:relate_2iters}, 
the optimal $D_{t}^*$ and $D_{t+1}^*$ correspond to the smallest
aggregate rate at iterations $t$ and $t+1$,  $R_t+R_{t+1}$.
Considering~\eqref{eq:DR}, we have
\begin{equation*}
R_{t}+R_{t+1}\!=\l( \!\frac{1}{2}\log_2\l(\frac{C_1}{D_{t}}\r)+\frac{1}{2}\log_2\l(\frac{C_1}{D_{t+1}}\r) \r) (1+o_t(1)).
\end{equation*}
Therefore, in the large $t$ limit, minimizing $R_{t}+R_{t+1}$ is
identical to maximizing the product $D_{t}D_{t+1}$.
Considering~\eqref{eq:iter_t}--\eqref{eq:relate_2iters}, our optimization problem becomes
maximization over $F(\widetilde{u}_{t+1})$, where
\begin{equation}\label{eq:optProb}
F(\widetilde{u}_{t+1}) =
(\widetilde{\sigma}_{t+2}^2-\widetilde{u}_{t+1})(\widetilde{g}_S^{-1}(\widetilde{u}_{t+1})-\widetilde{u}_t).
\end{equation}
Invoking Taylor's theorem~\eqref{eq:TaylorInverse} and considering that $C \in [-B, B]$, we solve the optimization problem~\eqref{eq:optProb} in two extremes: one with $C = B$ and the other with  $C=-B$. 

In the case of $C=B$, we obtain
\begin{equation*}
\begin{split}
F(\widetilde{u}_{t+1})=&-\frac{1}{\theta}\widetilde{u}_{t+1}^2+\frac{1}{\theta}\widetilde{u}_{t+1}\widetilde{\sigma}_{t+2}^2+B\widetilde{\sigma}_{t+2}^2\widetilde{u}_{t+1}^2-\\
&B\widetilde{u}_{t+1}^3-\widetilde{u}_t\widetilde{\sigma}_{t+2}^2+\widetilde{u}_t\widetilde{u}_{t+1}.
\end{split}
\end{equation*}
The maximum of $F(\widetilde{u}_{t+1})$ is achieved when $F'(\widetilde{u}_{t+1})=0$. That is,
\begin{equation}\label{eq:zeroDeriv}
F'(\widetilde{u}_{t+1})=-3B\widetilde{u}_{t+1}^2+\l(2B\widetilde{\sigma}_{t+2}^2-\frac{2}{\theta}\r)\widetilde{u}_{t+1}+\frac{\widetilde{\sigma}_{t+2}^2}{\theta}+\widetilde{u}_{t}=0.
\end{equation}
Considering  that $0<\widetilde{u}_{t+1}< \widetilde{u}_{t}$,
the root of the quadratic equation~\eqref{eq:zeroDeriv} is
\begin{equation}\label{eq:optSolution}
\widetilde{u}^*_{t+1}=\frac{1}{3B}\l[\l(B\widetilde{\sigma}_{t+2}^2-\frac{1}{\theta}\r)+A\r],
\end{equation}
where
\begin{equation}\label{eq:original_A}
A=\sqrt{\l(B\widetilde{\sigma}_{t+2}^2-\frac{1}{\theta}\r)^2+3B\l(\frac{\widetilde{\sigma}_{t+2}^2}{\theta}+\widetilde{u}_{t}\r)}.
\end{equation}
We can further simplify~\eqref{eq:original_A} as
\begin{equation}\label{eq:TaylorApprox}
\begin{split}
A=&\frac{1}{\theta}\sqrt{1+B(\theta \widetilde{\sigma}_{t+2}^2+B\theta^2 \widetilde{\sigma}_{t+2}^4+3\theta^2 \widetilde{u}_{t} )}\\
=&\frac{1}{\theta}\l[1+\frac{B}{2}(\theta \widetilde{\sigma}_{t+2}^2+B\theta^2\widetilde{\sigma}_{t+2}^4+
3\theta^2\widetilde{u}_{t})\r]+O(\widetilde{u}_t^2),
\end{split}
\end{equation}
Plugging~\eqref{eq:TaylorApprox} into~\eqref{eq:optSolution}, 
\begin{equation}\label{eq:opt_u_tp1}
\widetilde{u}_{t+1}^*=\frac{1}{2}(\widetilde{\sigma}_{t+2}^2+\theta \widetilde{u}_{t})+O(\widetilde{u}_{t}^2).
\end{equation}
Plugging~\eqref{eq:opt_u_tp1} into~\eqref{eq:iter_t} and~\eqref{eq:iter_tp1}, 
\begin{equation*}
\begin{split}
PD_t^*&=\frac{1}{2\theta}(\widetilde{\sigma}_{t+2}^2-\widetilde{u}_{t}\theta)+O(\widetilde{u}_t^2),\\
PD_{t+1}^*&=\frac{1}{2}(\widetilde{\sigma}_{t+2}^2-\widetilde{u}_t^2\theta )+O(\widetilde{u}_t^2),
\end{split}
\end{equation*}
which leads to
\begin{equation}\label{eq:optRatio_ori}
\frac{D_{t+1}^*}{D_t^*}=\theta(1+O(\widetilde{u}_t)).
\end{equation}

These steps provided the optimal relation between $D_{t}^*$ and $D_{t+1}^*$ when $C=B$.
For the other extreme case, $C=-B$,
similar steps will lead to~\eqref{eq:optRatio_ori},
where the differences between the results are higher order terms. Note that for any $C \in [-B, B]$ the higher order term is bounded between the two extremes.
Hence, the optimal $D_t^*$ and $D_{t+1}^*$
follow~\eqref{eq:optRatio_ori} leading to the first part of the claim~\eqref{eq:theorem1-1}.
Considering~\eqref{eq:DR} and~\eqref{eq:optRatio_ori},
\begin{equation*}
R_{t+1}^*-R_t^*=\frac{1}{2}\log_2 \l(\frac{1}{\theta}\r) (1+ o_t(1)).
\end{equation*}
Therefore, we obtain the second part of the claim~\eqref{eq:theorem1}.
\end{IEEEproof}

\subsection{Proof of Theorem~\ref{th:convergence}}\label{app:convergence}

\begin{IEEEproof}
Our proof is based on the assumption that lossy SE~\eqref{eq:SE_Q} holds.
Let us focus on an optimal coding rate sequence $\mathbf{R}^*=(R_1^*,\cdots,R_T^*)$.
Applying Taylor's theorem to calculate the ordinate of point $\widetilde{S}_{t+1}$ using its abscissa (Fig.~\ref{fig:evolveEMSE}), we obtain
\begin{equation}\label{eq:Taylor_u1}
\widetilde{u}_{t+1}^*=\theta(\widetilde{u}_t^*+PD_t^*)+O((\widetilde{u}_t^*)^2).
\end{equation}
Therefore,
\begin{equation}\label{eq:ratio1}
\frac{\widetilde{u}_{t+1}^*}{\widetilde{u}_t^*}=\theta+\frac{\theta PD_t^*}{\widetilde{u}_t^*}+O(\widetilde{u}_t^*).
\end{equation}
Similarly, we obtain
\begin{equation}\label{eq:ratio2}
\frac{\widetilde{u}_{t+2}^*}{\widetilde{u}_{t+1}^*}=\theta+\frac{\theta PD_{t+1}^*}{\widetilde{u}_{t+1}^*}+O(\widetilde{u}_t^*).
\end{equation}
Plugging~\eqref{eq:optRatio_ori} and~\eqref{eq:Taylor_u1} into~\eqref{eq:ratio2}, we obtain
\begin{equation}\label{eq:ratio2_new}
\begin{split}
\frac{\widetilde{u}_{t+2}^*}{\widetilde{u}_{t+1}^*}&=\theta+\frac{\theta PD_t^*(1+O(u_t^*))}{\widetilde{u}_t^*+PD_t^*+O((\widetilde{u}_t^*)^2)}+O(\widetilde{u}_t^*)\\
&=\theta+\frac{\theta PD_t^*}{\widetilde{u}_t^*+PD_t^*}+O(\widetilde{u}_t^*).
\end{split}
\end{equation}
On the other hand,  $\lim_{t\rightarrow\infty}\frac{\widetilde{u}_{t+1}^*}{\widetilde{u}_{t}^*}=\lim_{t\rightarrow\infty} \frac{\widetilde{u}_{t+2}^*}{\widetilde{u}_{t+1}^*}$.
Therefore, considering~\eqref{eq:ratio1} and~\eqref{eq:ratio2_new}, we obtain
\begin{equation*}
\lim_{t\rightarrow\infty} \frac{\theta PD_t^*}{\widetilde{u}_t^*} = \lim_{t\rightarrow\infty} \frac{\theta PD_t^*}{\widetilde{u}_t^*+PD_t^*},
\end{equation*}
which leads to $\lim_{t\rightarrow}\frac{D_t^*}{\widetilde{u}_t^*} =0$. We obtain~\eqref{eq:theorem2_2} by noting that the optimal EMSE at iteration t is $\epsilon_t^*=\frac{M}{N}\widetilde{u}_t^*$.
Plugging~\eqref{eq:theorem2_2} into~\eqref{eq:ratio1}, we obtain~\eqref{eq:theorem2_1}.
\end{IEEEproof}

\bibliography{cites}

\end{document}